\documentclass[12pt]{iopart}
\usepackage{iopams}
\usepackage{graphicx}
\usepackage[dvips]{color}
\usepackage{amssymb, amsfonts, amsbsy}

\begin{document}
\title[Cosmography: Extracting the Hubble series from the supernova data]    
{Cosmography:  Extracting the Hubble series from the supernova data}
\author{C\'eline Catto\"en and Matt Visser}
\address{School of Mathematics, Statistics, and Computer Science, \\
Victoria University of Wellington, PO Box 600, Wellington, New Zealand}
\ead{celine.cattoen@mcs.vuw.ac.nz, matt.visser@mcs.vuw.ac.nz}
\begin{abstract}

  Cosmography (cosmokinetics) is the part of cosmology that proceeds
  by making minimal dynamic assumptions.  One keeps the geometry and
  symmetries of FLRW spacetime, at least as a working hypothesis, but
  does not assume the Friedmann equations (Einstein equations), unless
  and until absolutely necessary.  By doing so it is possible to defer
  questions about the equation of state of the cosmological fluid, and
  concentrate more directly on the observational situation.  In
  particular, the ``big picture'' is best brought into focus by
  performing a fit of all available supernova data to the Hubble
  relation, from the current epoch at least back to redshift $z\approx
  1.75$.

  We perform a number of inter-related cosmographic fits to the {\sf
    legacy05} and {\sf gold06} supernova datasets. We pay particular
  attention to the influence of both statistical and systematic
  uncertainties, and also to the extent to which the choice of
  distance scale and manner of representing the redshift scale affect
  the cosmological parameters.  While the ``preponderance of
  evidence'' certainly suggests an accelerating universe, we would
  argue that (based on the supernova data) this conclusion is not
  currently supported ``beyond reasonable doubt''.  As part of the
  analysis we develop two particularly transparent graphical
  representations of the redshift-distance relation ---
  representations in which acceleration versus deceleration reduces to
  the question of whether the graph slopes up or down.

  Turning to the details of the cosmographic fits, three issues in
  particular concern us: First, the fitted value for the deceleration
  parameter changes significantly depending on whether one performs a
  $\chi^2$ fit to the luminosity distance, proper motion distance,
  angular diameter distance, or other suitable distance surrogate.
  Second, the fitted value for the deceleration parameter changes
  significantly depending on whether one uses the traditional redshift
  variable $z$, or what we shall argue is on theoretical grounds an
  improved parameterization $y=z/(1+z)$. Third, the published
  estimates for systematic uncertainties are sufficiently large that
  they certainly impact on, and to a large extent undermine, the usual
  purely statistical tests of significance.  We conclude that the case
  for an accelerating universe is considerably less watertight than
  commonly believed.

\vskip 0.250cm

\noindent
Based on a talk presented by Matt Visser at KADE 06, the ``Key
approaches to Dark Energy'' conference, Barcelona, August 2006;
follow up at GR18, Sydney, July 2007.

\vskip 0.1250cm


\vskip 0.1250cm
\noindent
arXiv:  gr-qc/0703122; Dated:  26 March 2007; \\
revised 10 May 2007; revised 31 July 2007; 
\LaTeX-ed \today.
  
\end{abstract}
\maketitle
\newtheorem{theorem}{Theorem}
\newtheorem{corollary}{Corollary}
\newtheorem{lemma}{Lemma}
\def\d{{\mathrm{d}}}
\def\implies{\Rightarrow}

\def\eg{{\it e.g.}}
\def\ie{{\it i.e.}}
\def\etc{{\it etc.}}
\def\sign{{\hbox{sign}}}
\def\eof{\Box}
\newenvironment{warning}{{\noindent\bf Warning: }}{\hfill $\eof$\break}
\markboth{Cosmography:  Extracting the Hubble series from the supernova data}{}
\clearpage

\section*{Comment on the revisions}

\begin{itemize}
\item 

In the version 2 revision  we have responded to community feedback by extending and clarifying the discussion, and by adding numerous additional references.  While our overall conclusions remain unchanged we have  rewritten our discussion of both statistical and systematic uncertainties  to use language that is more in line with the norms adopted by the supernova community.

\item

In particular we have adopted the nomenclature of the NIST reference on the ``Uncertainty of Measurement Results'' {\sf http://physics.nist.gov/cuu/Uncertainty/}, which is itself a distillation of NIST Technical Note 1297 ``Guidelines for Evaluating and Expressing the Uncertainty of NIST Measurement Results'',  which is in turn based on the ISO's ``Guide to the Expression of Uncertainty in Measurement'' (GUM). 

\item

This NIST document summarizes the norms and recommendations established by international agreement  between the NIST, the ISO, the BIPM, and the CIPM.

\item 

By closely adhering to this widely accepted standard, and to the norms  adopted by the supernova community, we hope we have now minimized the risk of miscommunication and misunderstanding.

\item

We emphasize: Our overall conclusions remain unchanged. The case for an accelerating universe is considerably less watertight than commonly believed.

\begin{itemize}
\item \emph{Regardless of one's views on how to combine formal estimates of uncertainty, the very fact that different distance scales yield data-fits with such widely discrepant values strongly suggests the need for extreme caution in interpreting the supernova data.}

\item \emph{Ultimately,  it is the fact that  figures \ref{F:ln_dQ_legacy05}--\ref{F:ln_dF_gold06}  do \emph{not} exhibit any overwhelmingly obvious trend that makes it so difficult to make a robust and reliable estimate of the sign of the deceleration parameter.}

\end{itemize}

\item
Version 3 now adds a little more discussion and historical context. Some historical graphs are added, plus some additional references, and a few clarifying comments. No physics changes.

\end{itemize}

\clearpage
\tableofcontents
\clearpage
\markboth{Cosmography:  Extracting the Hubble series from the supernova data}{}

\section{Introduction}

From various observations of the Hubble relation, most recently including the supernova data~\cite{legacy, legacy-url, gold, Riess2006a, Riess2006b, essence}, one is by now very accustomed to seeing many plots of luminosity distance $d_L$ versus redshift $z$. But are there better ways of representing the data?

For instance, consider cosmography (cosmokinetics) which is the part of cosmology
that proceeds by making minimal dynamic assumptions.
One keeps the geometry and symmetries of FLRW spacetime, 
\begin{equation}
\d s^2 = - c^2 \; \d t^2 + a(t)^2 \left\{ {\d r^2\over1-k\,r^2} + r^2 (\d\theta^2+\sin^2\theta\;\d\phi^2) \right\},
\end{equation}
at least as a working hypothesis,
but does not assume the Friedmann equations
(Einstein equations), unless and until absolutely necessary.
By doing so it is possible to defer questions about the
equation of state of the cosmological fluid, minimize the number of theoretical assumptions one is bringing to the table, and so concentrate more directly on the observational situation.

In particular, the ``big picture'' is best brought into focus by performing a global fit of all available supernova data to the Hubble relation, from the current epoch at least back to redshift $z\approx 1.75$. 
Indeed, all the discussion over acceleration versus deceleration, and the presence (or absence) of jerk (and snap) ultimately boils down, in a cosmographic setting, to doing a finite-polynomial truncated--Taylor series fit of the distance measurements (determined by supernovae and other means) to some suitable form of distance--redshift or distance--velocity relationship.  Phrasing the question to be investigated in this way keeps it as close as possible to Hubble's original statement of the problem, while minimizing the number of extraneous theoretical assumptions one is forced to adopt. For instance, it is quite standard to phrase the investigation in terms of the luminosity distance versus redshift relation~\cite{Weinberg, Peebles}:
\begin{eqnarray}
d_L(z) =  {c\; z\over H_0}
\Bigg\{ 1 + {1\over2}\left[1-q_0\right] {z} 
+ O(z^2) \Bigg\},
\label{E:Hubble1a}
\end{eqnarray}
and its higher-order extension~\cite{Chiba, Sahni, Jerk, Jerk2}
\begin{eqnarray}
\fl
d_L(z) =  {c\; z\over H_0}
\Bigg\{ 1 + {1\over2}\left[1-q_0\right] {z} 
-{1\over6}\left[1-q_0-3q_0^2+j_0+ {kc^2\over H_0^2\,a_0^2}\right] z^2
+ O(z^3) \Bigg\},
\label{E:Hubble1}
\end{eqnarray}
A central question thus has to do with the choice of the luminosity distance as the primary quantity of interest --- there are several other notions of cosmological distance that can be used, some of which (we shall see) lead to simpler and more tractable versions of the Hubble relation.
Furthermore, as will quickly be verified by looking at the derivation (see, for example,~\cite{Weinberg, Peebles, Chiba, Sahni, Jerk, Jerk2}, the standard Hubble law is actually a Taylor series expansion derived for small $z$, whereas much of the most interesting recent supernova data occurs at $z>1$. Should we even trust the usual formalism for large $z>1$?
Two distinct things could go wrong:
\begin{itemize}
\item The underlying Taylor series could fail to converge.
\item Finite truncations of the Taylor series might be a
               bad approximation to the exact result.
\end{itemize}
In fact, \emph{both} things happen. There are good mathematical and physical reasons for this undesirable behaviour, as we shall discuss below. We shall carefully explain just what goes wrong --- and suggest various ways of improving the situation. Our ultimate goal will be to find suitable forms of the Hubble relation that are well adapted to performing fits to all the available distance \emph{versus} redshift data.

Moreover --- once one stops to consider it carefully --- why should the cosmology community be so fixated on using the luminosity distance $d_L$ (or its logarithm, proportional to the distance modulus) and the redshift $z$ as the relevant parameters?
In principle, in place of luminosity distance $d_L(z)$ versus redshift $z$ one could just as easily plot 
$f(d_L,z)$ versus $g(z)$, choosing $f(d_L,z)$ and $g(z)$ to be arbitrary locally invertible functions, and \emph{exactly the same physics} would be encoded.
Suitably choosing the quantities to be plotted and fit will not change the physics, \emph{but it might improve statistical properties and insight}. (And we shall soon see that it will definitely improve the behaviour of the Taylor series.)

By comparing cosmological parameters obtained using multiple different fits of the Hubble relation to different distance scales and different parameterizations of the redshift we can then assess the robustness and reliability of the data fitting procedure.  In performing this analysis we had hoped to verify the robustness of the Hubble relation, and to possibly obtain improved estimates of cosmological parameters such as the deceleration parameter and jerk parameter, thereby complementing other recent cosmographic and cosmokinetic analyses such as~\cite{Blandford,    Blandford0, Shapiro, Caldwell, Elagory}, as well as other analyses that take a sometimes skeptical view of the totality of the observational data~\cite{paddy1, paddy2, paddy3, paddy4, barger}.  The actual results of our current cosmographic fits to the data are considerably more ambiguous than we had initially expected, and there are many subtle issues hiding in the simple phrase ``fitting the data''.  

In the following sections we first discuss the various cosmological distance scales, and the related versions of the Hubble relation. We then discuss technical problems with the usual redshift variable for $z>1$, and how to ameliorate them, leading to yet more versions of the Hubble relation. After discussing key features of the supernova data, we perform, analyze, and contrast  multiple fits to the Hubble relation --- providing discussions of model-building uncertainties (some technical details being relegated to the appendices) and sensitivity to systematic uncertainties.  Finally we present our results and conclusions: 
There is a disturbingly strong model-dependence in the resulting estimates for the deceleration parameter.  Furthermore, once realistic estimates of systematic uncertainties (based on the published data) are budgeted for it becomes clear that purely statistical estimates of goodness of fit are dangerously misleading.  While the ``preponderance of evidence'' certainly suggests an accelerating universe, we would argue that this conclusion is not currently supported ``beyond reasonable doubt'' ---  the supernova data (considered by itself) certainly \emph{suggests} an accelerating universe, it is not sufficient to allow us to reliably conclude that the universe is accelerating.\footnote{If one adds additional theoretical assumptions, such as by specifically fitting to a $\Lambda$-CDM model, the situation at first glance looks somewhat better ---  but this is then telling you as much about one's choice of theoretical model as it is about the observational situation.}

\section{Some history}

\begin{figure}[htb]
\begin{center}
\includegraphics[width=15cm]{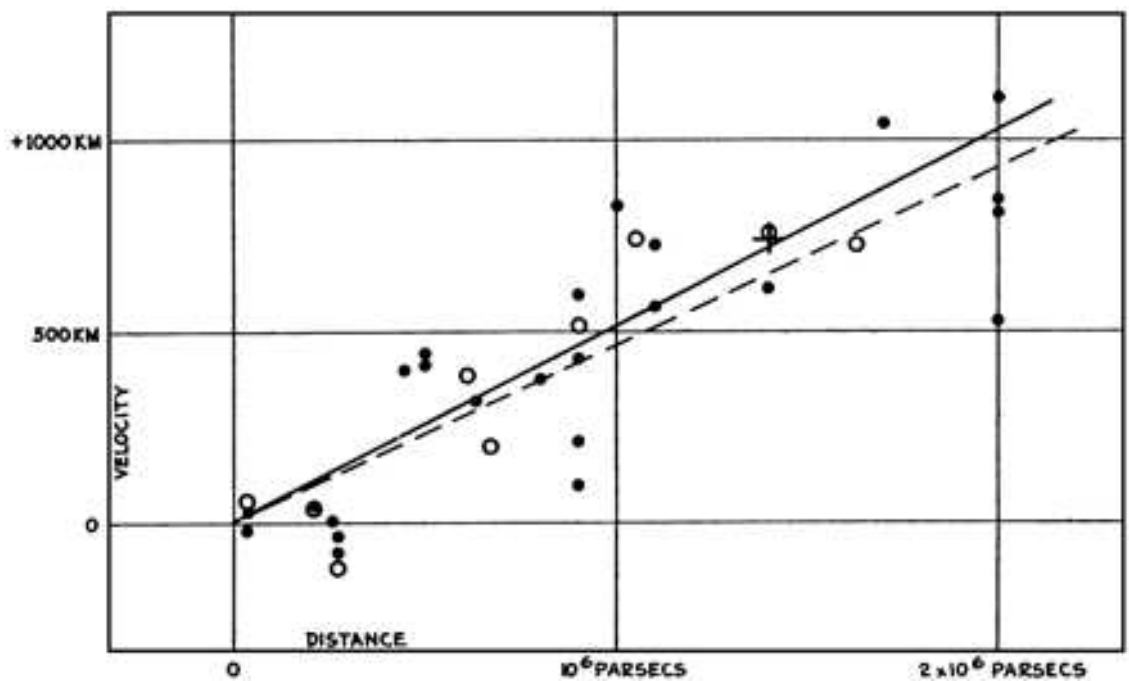}
\end{center}
\caption{\label{F:Hubble-1929}
Hubble's original 1929 plot~\cite{Hubble1929}. Note the rather large scatter in the data.}
\end{figure}

\begin{figure}[htb]
\begin{center}
\includegraphics[width=15cm]{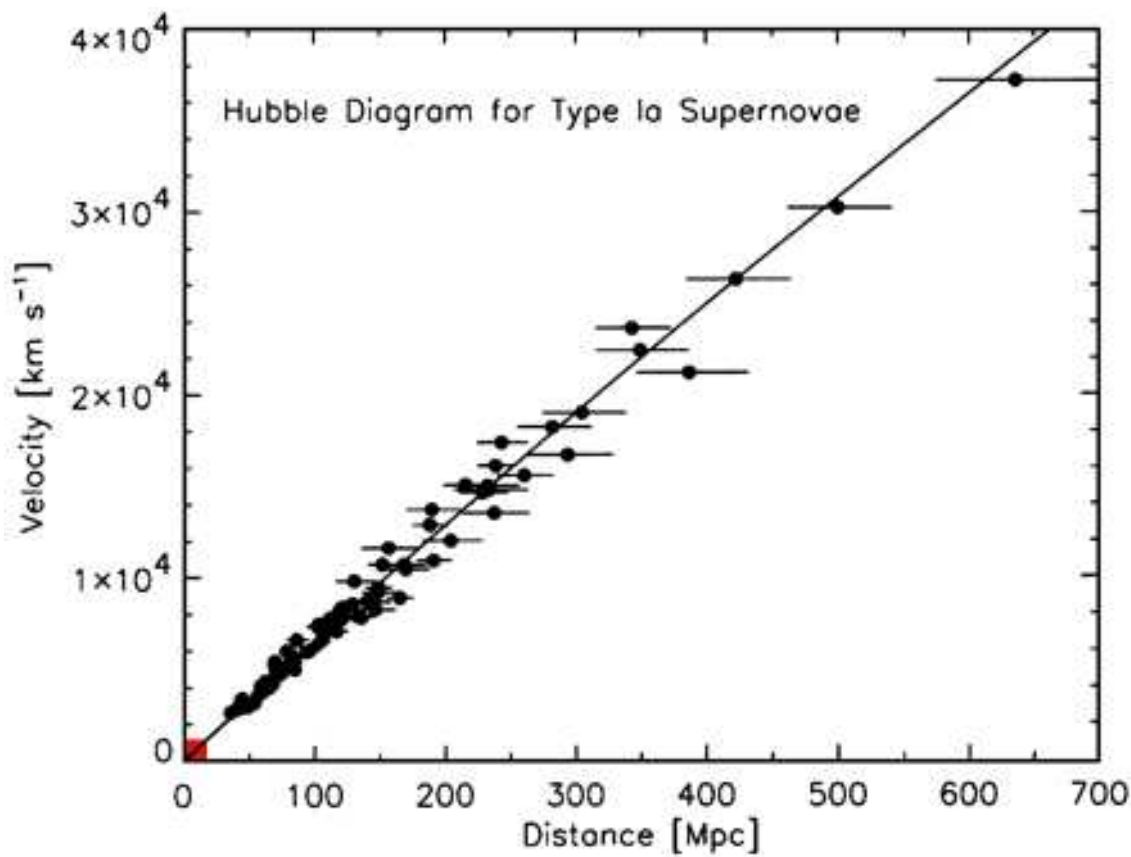}
\end{center}
\caption{\label{F:Hubble-2000}
Modern 2004 version of the Hubble plot.  From Kirshner~\cite{Kirshner}. The original 1929 Hubble plot is confined to the small red rectangle at the bottom left.}
\end{figure}

\begin{figure}[htb]
\begin{center}
\includegraphics[width=15cm]{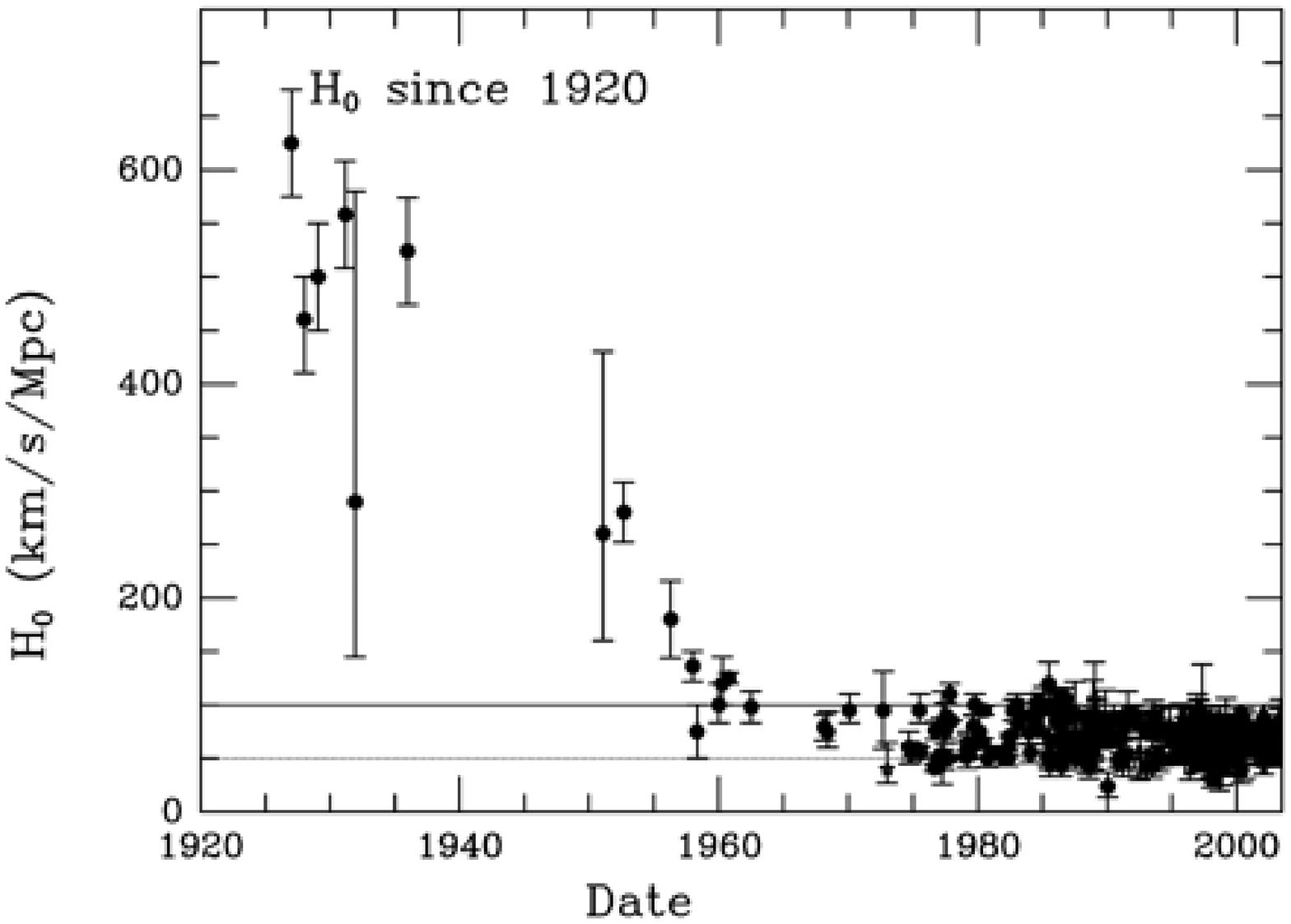}
\end{center}
\caption{\label{F:Hubble-variations}
Estimates of the Hubble parameter as a function of publication date.  From Kirshner~\cite{Kirshner}. Quote: ``At each epoch, the estimated error in the Hubble constant is small compared with the subsequent changes in its value. This result is a symptom of underestimated systematic errors.''}
\end{figure}

The need for a certain amount of caution in interpreting the observational data can clearly be inferred from a dispassionate reading of history. We reproduce below Hubble's original 1929 version of what is now called the Hubble plot (Figure~\ref{F:Hubble-1929})~\cite{Hubble1929}, a modern update from 2004 (Figure~\ref{F:Hubble-2000}) ~\cite{Kirshner}, and a very telling plot of the estimated value of the Hubble parameter \emph{as a function of publication date} (Figure~\ref{F:Hubble-variations})~\cite{Kirshner}. Regarding this last plot, Kirshner is moved to comment~\cite{Kirshner}:

\begin{quote}
``At each epoch, the estimated error in the Hubble constant is small compared with the subsequent changes in its value. This result is a symptom of underestimated systematic errors.''
\end{quote}

\begin{figure}[htb]
\begin{center}
\includegraphics[width=16cm]{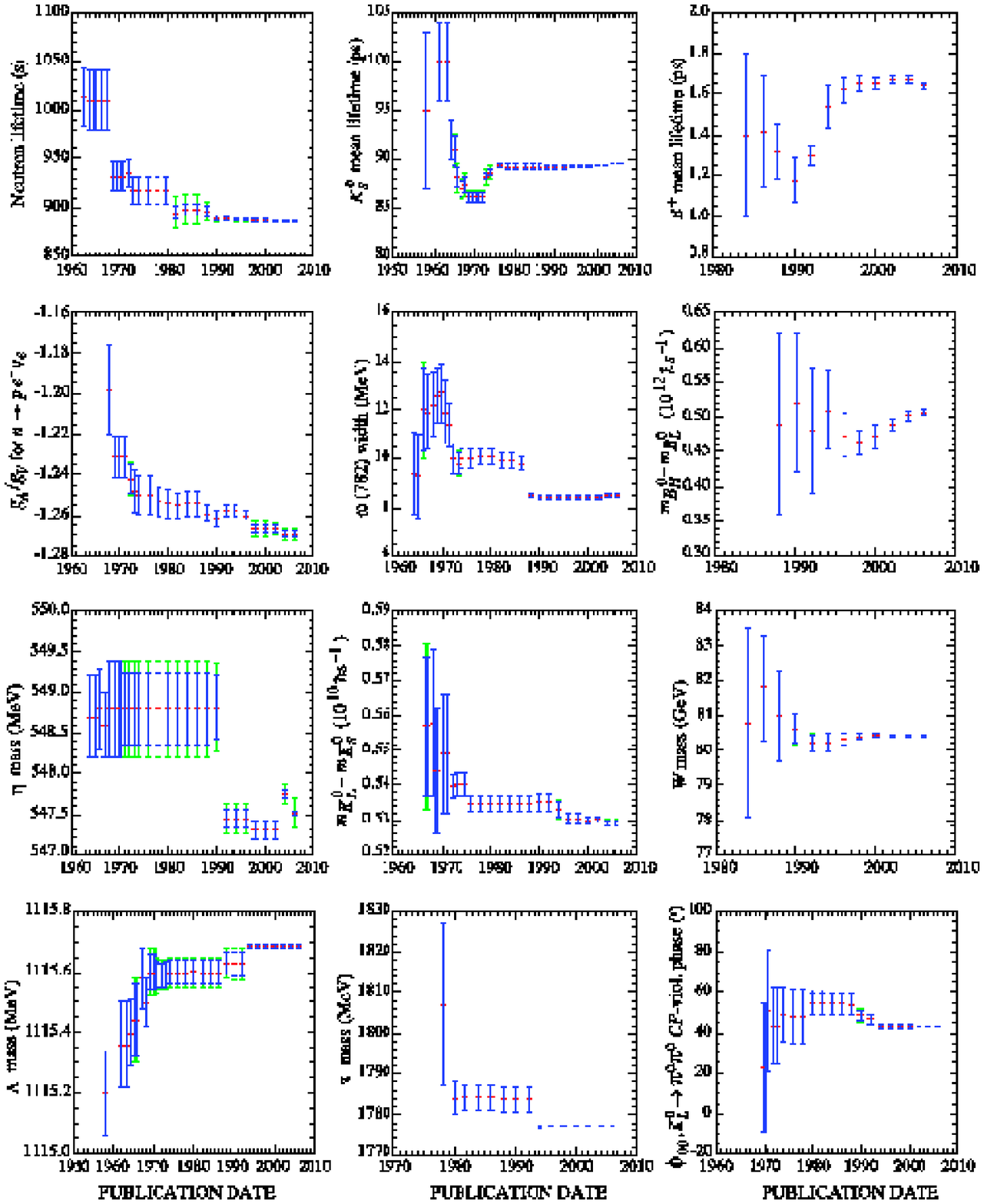}
\end{center}
\caption{\label{F:history}
Some historical plots of particle physics parameters as a function of publication date. From the Particle Data Group's 2006 Review of Particle Properties~\cite{PDG}. These plots strongly suggest that  the systematic under-estimating of systematic uncertainties is a generic phenomenon that cuts across disciplines and sub-fields, it is not a phenomenon that is limited to cosmology. }
\end{figure}

It is important to realise that the systematic under-estimating of systematic uncertainties is a generic phenomenon that cuts across disciplines and sub-fields, it is not a phenomenon that is limited to cosmology. For instance, the ``Particle Data Group'' [{\sf http://pdg.lbl.gov/}] in their bi-annual ``Review of Particle Properties'' publishes fascinating plots of estimated values of various particle physics parameters  \emph{as a function of publication date} (Figure~\ref{F:history})~\cite{PDG}.  These plots illustrate an aspect of the experimental and observational sciences that is often overlooked:

\begin{quote}
\emph{It is simply part of human nature to always  think the situation regarding systematic uncertainties is better than it actually is --- systematic uncertainties are systematically under-reported.}
\end{quote}

Apart from the many technical points we discuss in the body of the article below, ranging from the appropriate choice of cosmological distance scale, to the most appropriate version of redshift, to the  ``best'' way of representing the Hubble law, this historical perspective should also be kept in focus ---  ultimately the treatment of systematic uncertainties will prove to be an important component in estimating the reliability and robustness of the conclusions we can draw from the data.

\section{Cosmological distance scales}

In cosmology there are many different and equally natural definitions of the notion of   ``distance'' between two objects or events, whether directly observable or not.
For the vertical axis of the Hubble plot, instead of using the standard default choice of luminosity distance $d_L$, let  us now consider using one or more of:
\begin{itemize}

\item The ``photon flux distance'':
\begin{equation}
 d_F = {d_L\over(1+z)^{1/2}} .
\end{equation} 

\item The ``photon count distance'':
\begin{equation}
 d_P = {d_L\over(1+z)}.
\end{equation} 

\item The  ``deceleration distance'':
\begin{equation}
 d_Q = {d_L\over(1+z)^{3/2}}.
\end{equation} 

\item  The ``angular diameter distance'':
\begin{equation}
 d_A = {d_L\over(1+z)^2}.
\end{equation} 

\item The ``distance modulus'':
\begin{equation}
\mu_D = {5} \; \log_{10}[d_L/(10 \hbox{ pc})] = {5} \; \log_{10}[d_L/(1 \hbox{ Mpc})] +25.
\end{equation}

\item Or possibly some other surrogate for distance.
\end{itemize}
Some words of explanation and caution are in order here~\cite{Hogg}: 
\begin{itemize}

\item
The ``photon flux distance'' $d_F$ is based on the fact that it is often technologically easier to count the photon flux (photons/sec) than it is to bolometrically measure total energy flux (power) deposited in the detector. If we are counting photon number flux, rather than energy flux, then the photon number flux contains one fewer factor of $(1+z)^{-1}$. Converted to a distance estimator, the ``photon flux distance'' contains one extra factor of $(1+z)^{-1/2}$ as compared to the (power-based) luminosity distance.

\item 
The ``photon count distance''  $d_P$  is related to the total number of photons absorbed without regard to the rate at which they arrive. Thus the ``photon count distance'' contains one extra factor of $(1+z)^{-1}$ as compared to the (power-based) luminosity distance. Indeed D'Inverno~\cite{dInverno} uses what is effectively this photon count distance as his nonstandard definition for luminosity distance. Furthermore, though motivated very differently, this quantity is equal to Weinberg's definition of proper motion distance~\cite{Weinberg}, and is also equal to Peebles' version of angular diameter distance~\cite{Peebles}.  That is:
\begin{equation}
d_P= d_{L,\hbox{\scriptsize D'Inverno}}  = d_{\mathrm{proper},\mathrm{Weinberg}} = d_{A,\mathrm{Peebles}} .
\end{equation} 

\item 
The quantity $d_Q$ is (as far as we can tell)  a previously un-named quantity that seems to have no simple direct physical interpretation --- but we shall soon see why it is potentially useful, and why it is useful to refer to it as the ``deceleration distance''.

\item 
The quantity $d_A$ is Weinberg's definition of angular diameter distance~\cite{Weinberg}, corresponding to the physical size of the object \emph{when the light was emitted}, divided by its current angular diameter on the sky. This differs from Peebles' definition of angular diameter distance~\cite{Peebles}, which corresponds to what the size of the object would be at the current cosmological epoch if it had continued to co-move with the cosmological expansion (that is, the ``comoving size''), divided by its current angular diameter on the sky. Weinberg's $d_A$ exhibits the (at first sight perplexing, but physically correct) feature that beyond a certain point $d_A$ can actually \emph{decrease} as one moves to older objects that are clearly ``further'' away. In contrast Peebles' version of angular diameter distance is always increasing as one moves ``further'' away. Note that
\begin{equation}
d_{A,\mathrm{Peebles}}  = (1+z)\; d_A.
\end{equation}

\item 
Finally, note that the distance modulus can be rewritten in terms of traditional stellar magnitudes as 
\begin{equation}
\mu_D = \mu_\mathrm{apparent} - \mu_\mathrm{absolute}.
\end{equation}
The continued use of stellar magnitudes and the distance modulus in the context of cosmology is largely a matter of historical tradition, though we shall soon see that the logarithmic nature of the distance modulus has interesting and useful side effects. Note that we prefer as much as possible to deal with natural logarithms: $\ln x = \ln(10) \; \log_{10} x$. Indeed
\begin{equation}
\label{E:mu}
\mu_D = {5\over\ln10} \; \ln[d_L/(1 \hbox{ Mpc})] +25,
\end{equation}
so that
\begin{equation}
\label{E:d}
\ln[d_L/(1 \hbox{ Mpc})] = {\ln10\over5} [\mu_D - 25].
\end{equation}
\end{itemize}
Obviously
\begin{equation}
d_L \geq d_F \geq d_P \geq d_Q \geq d_A.
\end{equation} 
Furthermore these particular distance scales satisfy the property that they converge on each other, and converge on the naive Euclidean notion of distance, as $z\to0$.

To simplify subsequent formulae, it is now useful to define  the  ``Hubble distance''~\footnote{The ``Hubble distance'' $d_H = c/H_0$  is sometimes called the ``Hubble radius'', or the ``Hubble sphere", or even the ``speed of light sphere" [SLS]~\cite{Rothman}. Sometimes ``Hubble distance'' is used to refer to the naive estimate $d = d_H \; z$ coming from the linear part of the Hubble relation and ignoring all higher-order terms --- this is definitely \emph{not} our intended meaning.}
\begin{equation}
d_H = {c\over H_0},
\end{equation}
so that for $H_0 = 73\, {+ 3\atop- 4} \hbox{ (km/sec)/Mpc}$~\cite{PDG} we have
\begin{equation}
d_H = 4100\, {\textstyle {+ 240\atop- 160}} \hbox{ Mpc}.
\end{equation}
Furthermore we choose to set
\begin{equation}
\Omega_0=1+{kc^2\over H_0^2a_0^2} = 1 + {k \; d_H^2\over a_0^2}.
\end{equation}
For our purposes $\Omega_0$ is a purely cosmographic definition without dynamical content. (Only if one additionally invokes the Einstein equations in the form of the Friedmann equations does $\Omega_0$ have the standard interpretation as the ratio of total density to the Hubble density, but we would be prejudging things by making such an identification in the current cosmographic framework.) In the cosmographic framework $k/a_0^2$ is simply the present day curvature of space (not spacetime), while $d_H^{\;-2}=H_0^2/c^2$ is a measure of the contribution of expansion to the spacetime curvature of the FLRW geometry. More precisely, in a FRLW universe the Riemann tensor has (up to symmetry) only two non-trivial components. In an orthonormal basis:
\begin{equation}
R_{\hat\theta\hat\phi\hat\theta\hat\phi} = {k\over a^2} + {\dot a^2\over c^2 \; a^2}
= {k\over a^2} + {H^2\over c^2};
\end{equation}
\begin{equation}
R_{\hat t\hat r\hat t\hat r} = - {\ddot a\over c^2\; a } =  {q \; H^2\over c^2}.
\end{equation}
Then at arbitrary times $\Omega$ can be defined purely in terms of the Riemann tensor of the FLRW spacetime as   
\begin{equation}
\Omega = { R_{\hat\theta\hat\phi\hat\theta\hat\phi}(\dot a\to 0) 
           \over R_{\hat\theta\hat\phi\hat\theta\hat\phi} (k\to 0)}.
\end{equation}

\section{New versions of the Hubble law}

New versions of the Hubble law are easily calculated for each of these cosmological distance scales. 
Explicitly:
\begin{eqnarray}
\fl
d_L(z) =  {d_H \; z }
\Bigg\{ 1 - {1\over2}\left[-1+q_0\right] {z} 
+{1\over6}\left[q_0+3q_0^2-(j_0+\Omega_0)\right] z^2
+ O(z^3) \Bigg\}.
\end{eqnarray}

\begin{eqnarray}
\fl
d_F(z) =  {d_H \; z }
\Bigg\{ 1 - {1\over2}q_0 {z} 
+{1\over24}\left[3+10q_0+12q_0^2-4(j_0+\Omega_0)\right] z^2
+ O(z^3) \Bigg\}.
\end{eqnarray}

\begin{eqnarray}
\fl
d_P(z) =  {d_H \; z }
\Bigg\{ 1 - {1\over2}\left[1+q_0\right] {z} 
+{1\over6}\left[3+4q_0+3q_0^2-(j_0+\Omega_0)\right] z^2
+ O(z^3) \Bigg\}.
\end{eqnarray}

\begin{eqnarray}
\fl
d_Q(z) =  {d_H \; z }
\Bigg\{ 1   - {1\over2}\left[2+q_0\right] {z}
+{1\over24}\left[27+22q_0+12q_0^2-4(j_0+\Omega_0)\right] z^2
+ O(z^3) \Bigg\}.
\end{eqnarray}

\begin{eqnarray}
\fl
d_A(z) =  {d_H \; z }
\Bigg\{ 1 - {1\over2}\left[3+q_0\right] {z} 
+{1\over6}\left[12+7q_0+3q_0^2-(j_0+\Omega_0)\right] z^2
+ O(z^3) \Bigg\}.
\end{eqnarray}
If one simply wants to deduce (for instance)  the \emph{sign} of $q_0$, then it seems that plotting the ``photon flux distance'' $d_F$ versus $z$ would be a particularly good test --- simply check if the first nonlinear term in the Hubble relation curves up or down. 

In contrast, the Hubble law for the distance modulus itself is given by the more complicated expression
\begin{eqnarray}
\fl
\mu_D(z) &=&  25 + {5\over\ln(10)} \Bigg\{ \ln(d_H/\hbox{Mpc}) + \ln z 
\nonumber\\
\fl
&&
 + {1\over2}\left[1-q_0\right] {z} 
-{1\over24}\left[3-10q_0-9q_0^2+4(j_0+\Omega_0)\right] z^2
+ O(z^3) \Bigg\} .
\end{eqnarray}
However, when plotting $\mu_D$ versus $z$, most of the observed curvature in the plot comes from the universal ($\ln z$) term, and so carries no real information and is relatively uninteresting.  It is much better to rearrange the above as:
\begin{eqnarray}
\fl
\ln[d_L/(z \hbox{ Mpc})] &=& {\ln10\over5} [\mu_D - 25] - \ln z
\nonumber\\
\fl
&=& \ln(d_H/\hbox{Mpc})
\nonumber\\
\fl
&&
 - {1\over2}\left[-1+q_0\right] {z} 
+{1\over24}\left[-3+10q_0+9q_0^2-4(j_0+\Omega_0)\right] z^2
+ O(z^3).
\end{eqnarray}
In a similar manner one has
\begin{eqnarray}
\fl
\ln[d_F/(z \hbox{ Mpc})] &=& {\ln10\over5} [\mu_D - 25] - \ln z - {1\over2} \ln(1+z)
\nonumber\\
\fl
&=& \ln(d_H/\hbox{Mpc})
\nonumber\\
\fl
&&
 - {1\over2}q_0  {z} 
+{1\over24}\left[3+10q_0+9q_0^2-4(j_0+\Omega_0)\right] z^2
+ O(z^3).
\end{eqnarray}

\begin{eqnarray}
\fl
\ln[d_P/(z \hbox{ Mpc})] &=& {\ln10\over5} [\mu_D - 25] - \ln z -  \ln(1+z)
\nonumber\\
\fl
&=& \ln(d_H/\hbox{Mpc})
\nonumber\\
\fl
&&
 - {1\over2}\left[1+q_0\right]  {z} 
+{1\over24}\left[9+10q_0+9q_0^2-4(j_0+\Omega_0)\right] z^2
+ O(z^3).
\end{eqnarray}

\begin{eqnarray}
\fl
\ln[d_Q/(z \hbox{ Mpc})] &=& {\ln10\over5} [\mu_D - 25] - \ln z - {3\over2} \ln(1+z)
\nonumber\\
\fl
&=& \ln(d_H/\hbox{Mpc})
\nonumber\\
\fl
&&
 - {1\over2}\left[2+q_0\right]  {z} 
+{1\over24}\left[15+10q_0+9q_0^2-4(j_0+\Omega_0)\right] z^2
+ O(z^3).
\end{eqnarray}

\begin{eqnarray}
\fl
\ln[d_A/(z \hbox{ Mpc})] &=& {\ln10\over5} [\mu_D - 25] - \ln z - 2 \ln(1+z)
\nonumber\\
\fl
&=& \ln(d_H/\hbox{Mpc})
\nonumber\\
\fl
&&
 - {1\over2}\left[ 3 + q_0\right]  {z} 
+{1\over24}\left[21+10q_0+9q_0^2-4(j_0+\Omega_0)\right] z^2
+ O(z^3).
\end{eqnarray}
These logarithmic versions of the Hubble law have several advantages --- fits to these relations are easily calculated in terms of the observationally reported distance moduli $\mu_D$ and their estimated statistical uncertainties~\cite{legacy,  legacy-url, gold, Riess2006a, Riess2006b}.  (Specifically there is no need to transform the statistical uncertainties on the distance moduli beyond a universal multiplication by the factor $[\ln 10]/5$.) Furthermore the deceleration parameter $q_0$ is easy to extract as it has been ``untangled'' from both Hubble parameter and the combination ($j_0+\Omega_0$).

Note that it is always the combination ($j_0+\Omega_0$) that arises in these third-order Hubble relations, and that it is even in principle impossible to separately determine $j_0$ and $\Omega_0$ in a cosmographic framework.  The reason for this degeneracy is (or should be) well-known~\cite[p. 451]{Weinberg}: Consider the \emph{exact} expression for the luminosity distance in any FLRW universe, which is usually presented in the form~\cite{Weinberg, Peebles}
\begin{equation}
\label{E:exact}
d_L(z) = a_0 \; (1+z) \;
\sin_k \left\{ {c\over H_0 \, a_0} \int_0^z {H_0\over H(z)} \; \d z \right\},
\end{equation} 
where
\begin{equation}
\sin_k(x) = 
\left\{ 
\begin{array}{ll}
       \sin(x), & k=+1;\\
        x,       & k=0;\\
        \sinh(x), & k=-1.\\
\end{array} \right.
\end{equation} 
By inspection, even if one knows $H(z)$ exactly for all $z$ one cannot determine $d_L(z)$ without independent knowledge of $k$ and $a_0$. Conversely even if one knows $d_L(z)$ exactly for all $z$ one cannot determine $H(z)$ without independent knowledge of $k$ and $a_0$. Indeed let us rewrite this exact result in a slightly different fashion as
\begin{equation}
d_L(z) = a_0 \; (1+z) \; 
{\sin\left\{ {\displaystyle {\sqrt{k} \, d_H\over a_0} \; \int_0^z {H_0\over H(z)}\;\d z } \right\} 
\over 
\sqrt{k} },
\end{equation}
where this result now holds for all $k$ provided we interpret the $k=0$ case in the obvious limiting fashion.
Equivalently, using the cosmographic $\Omega_0$ as defined above we have the \emph{exact} cosmographic result that for all $\Omega_0$:
\begin{equation}
d_L(z) = d_H \; (1+z) \; {\sin\left\{ \sqrt{\Omega_0-1} \;
{\displaystyle \int_0^z  {H_0\over H(z)}} \;\d z \right\} \over \sqrt{\Omega_0-1} }.
\end{equation}
This form of the exact Hubble relation makes it clear that an independent determination of $\Omega_0$ (equivalently, $k/a_0^2$), is needed to complete the link between $a(t)$ and $d_L(z)$. When Taylor expanded in terms of $z$, this expression leads to a degeneracy at third-order, which is  where $\Omega_0$ [equivalently $k/a_0^2$] first enters into the Hubble series~\cite{Jerk, Jerk2}.

What message should we take from this discussion? There are many physically equivalent versions of the Hubble law, corresponding to many slightly different physically reasonable definitions of distance, and whether we choose to present the Hubble law linearly or logarithmically.  If one were to have arbitrarily small scatter/error bars on the observational data, then the choice of which Hubble law one chooses to fit to would not matter. In the presence of significant scatter/uncertainty there is a risk that the fit might depend strongly on the choice of Hubble law one chooses to work with. (And if the resulting values of the deceleration parameter one obtains do depend significantly on which distance scale one uses, this is evidence that one should be very cautious in interpreting the results.)
Note that the two versions of the Hubble law based on ``photon flux distance''  $d_F$ stand out in terms of making the deceleration parameter  easy to visualize and extract.

\section{Why is the redshift expansion badly behaved for $z>1$?}

In addition to the question of which distance measure one chooses to use, there is a basic and fundamental physical and mathematical reason why the traditional redshift expansion breaks down for $z>1$.

\subsection{Convergence}

Consider the exact Hubble relation (\ref{E:exact}). 
This is certainly nicely behaved, and possesses no obvious poles or singularities, (except possibly at a  turnaround event where $H(z)\to0$, more on this below).
However if we attempt to develop a Taylor series expansion in redshift $z$, using what amounts to the \emph{definition} of the Hubble $H_0$, deceleration $q_0$,
 and jerk $j_0$ parameters, then:
\begin{equation}
\fl
{1\over 1+z} =  {a(t)\over a_0} = 1 + H_0 \; (t-t_0) -  {q_0 \; H_0^2\over 2!}  \;(t-t_0)^2 
+{j_0\; H_0^3\over 3!} \;(t-t_0)^3
+ O([t-t_0]^4).
\end{equation} 
Now this particular Taylor expansion manifestly has a pole at $z=-1$, corresponding to the instant (either  at finite or infinite time) when the universe has expanded to infinite volume, $a=\infty$. Note that a \emph{negative} value for $z$  corresponds to $a(t)>a_0$, that is: In an expanding universe $z<0$ corresponds to the \emph{future}.
Since there is an explicit pole at $z=-1$, by standard complex variable theory the radius of convergence is \emph{at most} $|z|=1$, so that this series \emph{also} fails to converge for $z > 1$, when the universe was less than half its current size. 

Consequently when reverting this power series to obtain lookback time $T=t_0-t$ as a function $T(z)$ of $z$, we should not expect that series to converge for $z >1$.  Ultimately, when written in terms of $a_0$, $H_0$, $q_0$, $j_0$, and a power series expansion in redshift $z$ you should not expect $d_L(z)$ to converge for $z > 1$.

\begin{figure}[htb]
\begin{center}
\input{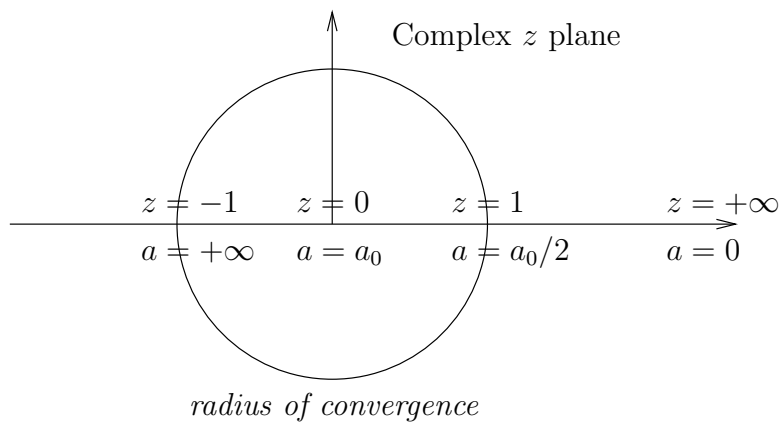}
\end{center}
\caption{\label{F:convergence_radius_z}
Qualitative sketch of the behaviour of the scale factor $a$ and the radius of convergence of the Taylor series in $z$-redshift.

}
\end{figure}

Note that the \emph{mathematics} that goes into this result is that the radius of convergence of any power series is the distance to the closest singularity in the complex plane, while the relevant \emph{physics} lies in the fact that on physical grounds we should not expect to be able to extrapolate forwards  beyond $a=\infty$, corresponding to $z=-1$. Physically we should expect this argument to hold for any observable quantity when expressed as a function of redshift and Taylor expanded around $z=0$ --- the radius of convergence of the Taylor series must be less than or equal to unity. (Note that the radius of convergence might actually be \emph{less} than unity, this occurs if some other singularity in the complex $z$ plane is closer than the breakdown in predictability associated with attempting to drive $a(t)$ ``past'' infinite expansion, $a=\infty$.) Figure \ref{F:convergence_radius_z} illustrates the radius of convergence in the complex plane of the Taylor series expansion in terms of $z$.

Consequently, we must conclude that observational data regarding $d_L(z)$ for $z > 1$ is not going to be particularly useful in fitting $a_0$, $H_0$, $q_0$, and $j_0$, to the usual \emph{traditional} version of the Hubble relation.

\subsection{Pivoting}
A trick that is sometimes used to improve the behaviour of the Hubble law is to Taylor expand around some nonzero value of $z$, which might be called the ``pivot''. That is, we take
\begin{equation}
 z = z_{pivot}+\Delta z,
\end{equation} 
and expand in powers of $\Delta z$. If we choose to do so, then observe
\begin{equation}
\fl
{1\over 1+z_{pivot}+\Delta z} =  1 + H_0 \; (t-t_0) - {1\over2} \; q_0 \; H_0^2 \;(t-t_0)^2 
+{1\over3!}\;  j_0\; H_0^3 \;(t-t_0)^3 
+ O([t-t_0]^4).\;\;
\end{equation} 
The pole is now located at:
\begin{equation}
\Delta z = -(1+z_{pivot}),
\end{equation} 
which again physically corresponds to a universe that has undergone infinite expansion, $a=\infty$. The radius of convergence is now 
\begin{equation}
|\Delta z| \leq (1+z_{pivot}),
\end{equation}
and we expect the pivoted version of the Hubble law to fail for
\begin{equation}
z > 1 + 2 \; z_{pivot}.
\end{equation} 
So pivoting is certainly helpful, and can in principle extend the convergent region of the Taylor expanded Hubble relation to somewhat higher values of $z$, but maybe we can do even better?

\subsection{Other singularities}
Other singularities that might further restrict the radius of convergence of  the Taylor expanded Hubble law (or any other Taylor expanded physical observable) are also important. Chief among them are the singularities (in the Taylor expansion) induced by turnaround events. If the universe has a minimum scale factor $a_\mathrm{min}$ (corresponding to a ``bounce'') then clearly it is meaningless to expand beyond
\begin{equation}
1+z_\mathrm{max} = a_0/a_\mathrm{min};  \qquad z_\mathrm{max} = a_0/a_\mathrm{min}-1;
\end{equation}
implying that we should restrict our attention to the region
\begin{equation}
|z| <  z_\mathrm{max} = a_0/a_\mathrm{min}-1.
\end{equation}
Since for other reasons we had already decided we should restrict attention to $|z|<1$, and since on observational grounds we certainly expect any ``bounce'', if it occurs at all, to occur for $z_\mathrm{max}\gg 1$, this condition provides no new information. 

On the other hand, if the universe has a moment of maximum expansion, and then begins to recollapse, then it is meaningless to extrapolate beyond
\begin{equation}
1+z_\mathrm{min} = a_0/a_\mathrm{max};  \qquad z_\mathrm{min} = -[1-a_0/a_\mathrm{max}];
\end{equation}
implying that we should restrict our attention to the region
\begin{equation}
|z| <  1 - a_0/a_\mathrm{max}.
\end{equation}
This relation now does provide us with additional constraint, though (compared to the $|z|<1$ condition) the bound is not appreciably tighter unless we are ``close" to a point of maximum expansion. Other singularities could lead to additional constraints.

\section{Improved redshift variable for the Hubble relation}

Now it must be admitted that the traditional redshift has a particularly simple physical interpretation:
\begin{equation}
 1+z = {\lambda_0\over\lambda_e} = {a(t_0)\over a(t_e)},
\end{equation} 
so that
\begin{equation}
z = {\lambda_0-\lambda_e\over\lambda_e} = {\Delta\lambda\over \lambda_e}.
\end{equation} 
That is, $z$ is the change in wavelength divided by the \emph{emitted} wavelength.
This is certainly simple, but there's at least one other \emph{equally simple} choice. Why not use:
\begin{equation}
y = {\lambda_0-\lambda_e\over\lambda_0} = {\Delta\lambda\over \lambda_0}\;?
\end{equation} 
That is, define $y$ to be the change in wavelength divided by the \emph{observed} wavelength.
This implies
\begin{equation}
 1-y = {\lambda_e\over\lambda_0} = {a(t_e)\over a(t_0)} = {1\over1+z}.
\end{equation} 
Now similar expansion variables have certainly been considered before. (See, for example,  Chevalier and   Polarski~\cite{Polarski}, who effectively worked with the dimensionless quantity $b=a(t)/a_0$, so that $y=1-b$. Similar ideas have also appeared in several related works~\cite{Linder, Linder2, Bassett, Martin}. Note that these authors have typically been interested in parameterizing the so-called $w$-parameter, rather than specifically addressing the Hubble relation.)

Indeed, the variable $y$ introduced above has some very nice properties:
\begin{equation}
y = {z\over1+z}; \qquad z={y\over1-y}.
\end{equation} 
In the past (of an expanding universe)
\begin{equation}
z \in (0,\infty); \qquad  y\in (0,1);
\end{equation}
while in the future
\begin{equation}
z \in (-1,0); \qquad  y\in (-\infty,0).
\end{equation}
So the variable $y$ is both easy to compute, and when extrapolating back to the Big Bang has a nice finite range $(0,1)$.
We will refer to this variable as the \emph{$y$-redshift}.
(Originally when developing these ideas we had intended to use the variable $y$ to develop orthogonal polynomial expansions on the finite interval $y\in [0,1]$. This is certainly possible, but we shall soon see that given the current data, this is somewhat overkill, and simple polynomial fits in $y$ are adequate for our purposes.)

In terms of the variable $y$ it is easy to extract a new version of the Hubble law by simple substitution:
\begin{eqnarray} \label{dL}
\fl
d_L(y) =  {d_H\; y}
\Bigg\{ 1 - {1\over2}\left[-3+q_0\right] {y} 
+{1\over6}\left[12-5q_0+3q_0^2-(j_0+ \Omega_0) \right] y^2
+ O(y^3) \Bigg\}.
\end{eqnarray}
This still looks rather messy, in fact as messy as before  --- one might justifiably ask in what sense is this new variable any real improvement?

\begin{figure}[htb]
\begin{center}
\input{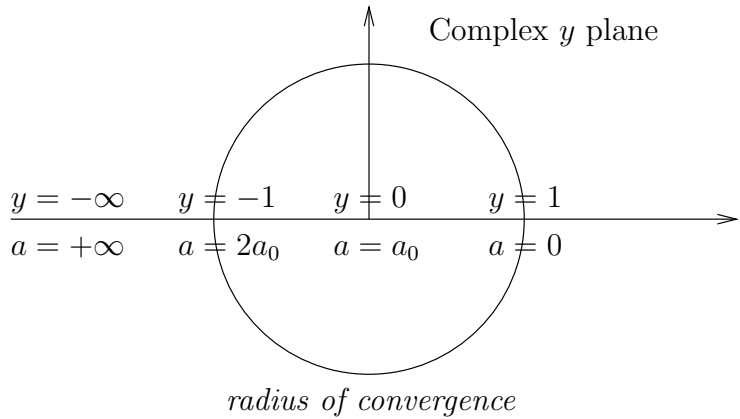}
\end{center}
\caption{\label{F:convergence_radius_y}
Qualitative sketch of the behaviour of the scale factor $a$ and the radius of convergence of the Taylor series in $y$-redshift.
}
\end{figure}

First, when expanded in terms of $y$, the formal radius of convergence covers much more of the physically interesting region. Consider:
\begin{eqnarray}
\fl
1-y =  1 + H_0 \; (t-t_0) - {1\over2} \; q_0 \; H_0^2 \;(t-t_0)^2 
+{1\over3!}\;  j_0\; H_0^3 \;(t-t_0)^3
+ O([t-t_0]^4).
\end{eqnarray}
This expression now has no poles, so upon reversion of the series lookback time $T=t_0-t$ should be well behaved as a function $T(y)$ of $y$ --- at least all the way back to the Big Bang.
(We now expect, on physical grounds, that the power series is likely to break down if one tries to extrapolate backwards \emph{through} the Big Bang.) Based on this, we now expect $d_L(y)$, as long as it is expressed as a Taylor series in the variable $y$, to be a well-behaved power series
all the way to the Big Bang. 
In fact, since
\begin{equation}
 y = +1    \qquad \Leftrightarrow  \qquad    \hbox{Big Bang},
\end{equation} 
we expect the radius of convergence to be given by $|y|=1$, so that the series converges for 
\begin{equation}
 |y| < 1.
\end{equation} 
Consequently, when looking into the future, in terms of the variable $y$ we expect to encounter problems at $y = -1$, when the universe
    has expanded to twice its current size.  Figure \ref{F:convergence_radius_y} illustrates the radius of convergence in the complex plane of the Taylor series expansion in terms of $y$.

Note the tradeoff here --- $z$ is a useful expansion parameter for arbitrarily large universes, but breaks down for a universe half its current size or less; in contrast $y$ is a useful expansion parameter all the way back to the Big Bang, but breaks down for a universe double its current size or more. Whether or not $y$ is more suitable than $z$ depends very much on what you are interested in doing. This is illustrated in Figures \ref{F:convergence_radius_z} and \ref{F:convergence_radius_y}. For the purposes of this article we are interested in high-redshift supernovae --- and we want to probe rather early times --- so it is definitely $y$ that is more appropriate here. Indeed the furthest supernova for which we presently have both spectroscopic data and an estimate of the distance occurs at $z=1.755$~\cite{Riess2006a},  corresponding to $y=0.6370$.
Furthermore, using the variable $y$ it is easier to plot very large redshift datapoints. For example, (though we shall not pursue this point in this article), the Cosmological Microwave Background is located at $z_\mathrm{CMB}=1088$, which corresponds to $y_\mathrm{CMB}=0.999$. This point is not ``out of range'' as it would be if one uses the variable $z$.

\section{More versions of the Hubble law}

In terms of this new redshift variable, the ``linear in distance'' Hubble relations are:
\begin{eqnarray}
\fl
d_L(y) =  {d_H\; y}
\Bigg\{ 1 - {1\over2}\left[-3+q_0\right] {y} 
+{1\over6}\left[12-5q_0+3q_0^2-(j_0+\Omega_0) \right] y^2
+ O(y^3) \Bigg\}.
\end{eqnarray}

\begin{eqnarray}
\fl
d_F(y) =  {d_H\; y}
\Bigg\{ 1 - {1\over2}\left[-2+q_0\right] {y} 
+{1\over24}\left[27-14q_0+12q_0^2-4(j_0+\Omega_0)\right] y^2
+ O(y^3) \Bigg\}.
\end{eqnarray}

\begin{eqnarray}
\fl
d_P(y) =  {d_H\; y}
\Bigg\{ 1 - {1\over2}\left[-1+q_0\right] {y} 
+{1\over6}\left[3-2q_0+3q_0^2-(j_0+ \Omega_0)\right] y^2
+ O(y^3) \Bigg\}.
\end{eqnarray}

\begin{eqnarray}
\fl
d_Q(y) =  {d_H\; y}
\Bigg\{ 1 - {q_0\over2} {y} 
+{1\over12}\left[3-2q_0+12q_0^2-4(j_0+ \Omega_0)\right] y^2
+ O(y^3) \Bigg\}.
\end{eqnarray}

\begin{eqnarray}
\fl
d_A(y) =  {d_H\; y}
\Bigg\{ 1 - {1\over2}\left[1+q_0\right] {y} 
+{1\over6}\left[q_0+3q_0^2-(j_0+\Omega_0)\right] y^2
+ O(y^3) \Bigg\}.
\end{eqnarray}
Note that in terms of the $y$ variable it is the ``deceleration distance'' $d_Q$ that has the deceleration parameter $q_0$ appearing in the simplest manner. Similarly, the ``logarithmic in distance'' Hubble relations are:
\begin{eqnarray}
\fl
\ln[d_L/(y \hbox{ Mpc})] &=& {\ln10\over5} [\mu_D - 25] - \ln y
\nonumber\\
\fl
&=& \ln(d_H/\hbox{Mpc})
\nonumber\\
\fl
&&
 -{1\over2}\left[-3+q_0\right] {y} 
+{1\over24}\left[21-2q_0+9q_0^2-4(j_0+\Omega_0)\right] y^2
+ O(y^3).
\end{eqnarray}

\begin{eqnarray}
\fl
\ln[d_F/(y \hbox{ Mpc})] &=& {\ln10\over5} [\mu_D - 25] - \ln y + {1\over2} \ln(1-y)
\nonumber\\
\fl
&=& \ln(d_H/\hbox{Mpc})
\nonumber\\
\fl
&&
 - {1\over2}\left[-2+q_0\right]  {y} 
+{1\over24}\left[15-2q_0+9q_0^2-4(j_0+\Omega_0)\right] y^2
+ O(y^3).
\end{eqnarray}

\begin{eqnarray}
\fl
\ln[d_P/(y \hbox{ Mpc})] &=& {\ln10\over5} [\mu_D - 25] - \ln y +  \ln(1-y)
\nonumber\\
\fl
&=& \ln(d_H/\hbox{Mpc})
\nonumber\\
\fl
&&
 - {1\over2}\left[-1+q_0\right]  {y} 
+{1\over24}\left[9-2q_0+9q_0^2-4(j_0+\Omega_0)\right] y^2
+ O(y^3).
\end{eqnarray}

\begin{eqnarray}
\fl
\ln[d_Q/(y \hbox{ Mpc})] &=& {\ln10\over5} [\mu_D - 25] - \ln y + {3\over2} \ln(1-y)
\nonumber\\
\fl
&=& \ln(d_H/\hbox{Mpc})
\nonumber\\
\fl
&&
 - {1\over2}q_0\,  {y} 
+{1\over24}\left[3-2q_0+9q_0^2-4(j_0+\Omega_0)\right] y^2
+ O(y^3).
\end{eqnarray}

\begin{eqnarray}
\fl
\ln[d_A/(y \hbox{ Mpc})] &=& {\ln10\over5} [\mu_D - 25] - \ln y + 2 \ln(1-y)
\nonumber\\
\fl
&=& \ln(d_H/\hbox{Mpc})
\nonumber\\
\fl
&&
 - {1\over2}\left[ 1 + q_0\right]  {y} 
+{1\over24}\left[-3-2q_0+9q_0^2-4(j_0+\Omega_0)\right] y^2
+ O(y^3).
\end{eqnarray}
Again note that the ``logarithmic in distance'' versions of the Hubble law are attractive in terms of maximizing the disentangling between Hubble distance, deceleration parameter, and jerk.
Now having a selection of Hubble laws on hand, we can start to confront the observational data to see what it is capable of telling us.

\section{Supernova data}

For the plots below we have used data from the supernova legacy survey ({\sf legacy05})~\cite{legacy,legacy-url} and the Riess \emph{et.~al.}~``gold'' dataset of 2006 ({\sf gold06})~\cite{Riess2006a}. 

\subsection{The {\sf legacy05} dataset}

The data is available in published form~\cite{legacy}, and in a slightly different format, via internet~\cite{legacy-url}. (The differences amount to minor matters of choice in the presentation.) The final processed result reported for each 115 of the supernovae is a redshift $z$, a luminosity modulus $\mu_B$, and an uncertainty in the luminosity modulus. The luminosity modulus can be converted into a luminosity distance via the formula
\begin{equation}
 d_L = (1 \hbox{ Megaparsec}) \times10^{(\,\mu_B+\mu_\mathrm{offset}-25)/5}.
\end{equation}
The reason for the ``offset'' is that supernovae by themselves only determine the \emph{shape} of the Hubble relation (\ie, $q_0$, $j_0$, \etc), but not its absolute \emph{slope} (\ie, $H_0$) --- this is ultimately due to the fact that we do not have good control of the absolute luminosity of the supernovae in question. The offset $\mu_\mathrm{offset}$ can be  chosen to match the known value of $H_0$ coming from other sources. (In fact the data reported in the published article~\cite{legacy} has already been normalized in this way to the ``standard value'' $H_{70} = 70 \hbox{ (km/sec)/Mpc}$, corresponding to Hubble distance $d_{70}=c/H_{70} =  4283  \hbox{ Mpc}$, whereas the data available on the website~\cite{legacy-url} has \emph{not} been normalized in this way --- which is why $\mu_B$ as reported on the website is systematically $19.308$ stellar magnitudes smaller than that in the published article.) 

The other item one should be aware of concerns the error bars: The error bars reported in the published article~\cite{legacy} are photometric uncertainties only --- there is an additional source of error to do with the intrinsic variability of the supernovae.  In fact, if you take the \emph{photometric} error bars seriously as estimates of the \emph{total} uncertainty, you would have to reject the hypothesis that we live in a standard FLRW universe.  Instead, intrinsic variability in the supernovae is by far the most widely accepted interpetation.  Basically one uses the  ``nearby'' dataset to estimate an
 intrinsic variability that makes chi-squared look reasonable.
This intrinsic variability of 0.13104 stellar magnitudes~\cite{legacy-url,Blandford}) has been estimated by looking at low redshift supernovae (where we have good measures of absolute distance from other techniques), and has been included in the error bars reported on the website~\cite{legacy-url}.
Indeed
\begin{equation} 
\label{total_error}
(\hbox{uncertainty})_\mathrm{website} = 
\sqrt{ (\hbox{intrinsic variability})^2 + 
(\hbox{uncertainty})_\mathrm{article}^2 }. 
\end{equation} 
With these key features of the supernovae data kept in mind, conversion to luminosity distance and estimation of scientifically reasonable error bars (suitable for chi-square analysis) is straightforward.

\begin{figure}[htb]
\begin{center}
\includegraphics[width=12cm]{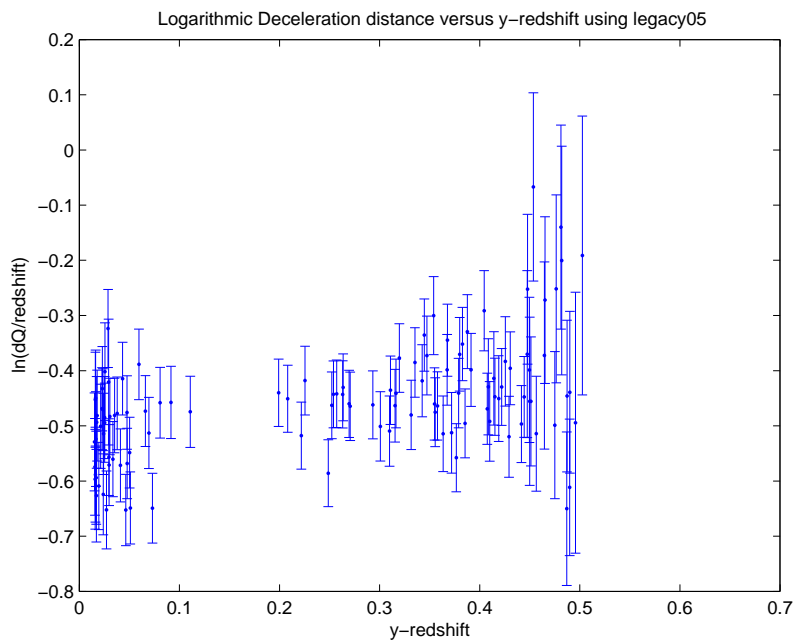}
\end{center}
\caption{\label{F:ln_dQ_legacy05}
The normalized logarithm of the deceleration distance, $\ln(d_Q/[y \hbox{ Mpc}])$, as a function of the $y$-redshift using the {\sf legacy05} dataset~\cite{legacy, legacy-url}.}
\end{figure}

\begin{figure}[htb]
\begin{center}
\includegraphics[width=12cm]{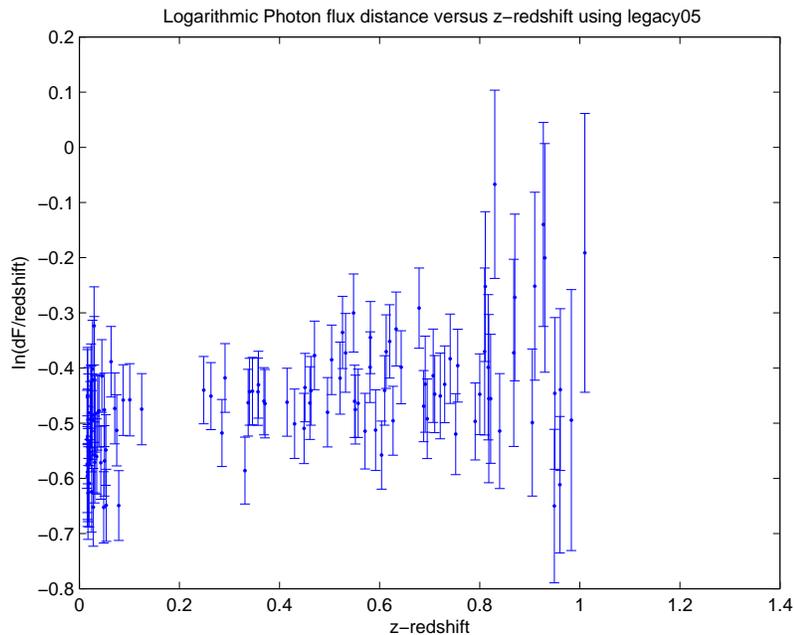}
\end{center}
\caption{\label{F:ln_dF_legacy05}
The normalized logarithm of the photon flux distance, $\ln(d_F/[z \hbox{ Mpc}])$, as a function of the $z$-redshift using the {\sf legacy05} dataset~\cite{legacy, legacy-url}.}
\end{figure}

To orient oneself, figure  \ref{F:ln_dQ_legacy05}  focuses on the deceleration distance $d_Q(y)$, and plots $\ln(d_Q/[ y \hbox{ Mpc}])$ versus $y$. Visually, the curve appears close to flat, at least out to $y\approx 0.4$, which is an unexpected oddity that merits further investigation --- since it seems to imply an ``eyeball estimate'' that $q_0\approx 0$.   Note that this is not a plot of ``statistical residuals'' obtained after curve fitting --- rather this can be interpreted as a plot of ``theoretical residuals'', obtained by first splitting off the linear part of the Hubble law (which is now encoded in the intercept with the vertical axis), and secondly choosing the quantity to be plotted so as to make the slope of the curve at zero particularly easy to interpret in terms of the deceleration parameter. The fact that there is considerable ``scatter'' in the plot should not be thought of as an artifact due to a ``bad'' choice of variables --- instead this choice of variables should be thought of as ``good'' in the sense that they provide an honest basis for dispassionately assessing the quality of the  data that currently goes into determining the deceleration parameter. Similarly, figure  \ref{F:ln_dF_legacy05}  focuses on the photon flux distance $d_F(z)$,  and plots $\ln(d_F/[z \hbox{ Mpc}])$ versus $z$. Visually, this curve is again very close to flat, at least out to $z\approx 0.4$.  This again gives one a feel for just how tricky it is to reliably estimate the deceleration parameter $q_0$ from the data.

\subsection{The {\sf gold06} dataset}

Our second collection of data is the {\sf gold06} dataset~\cite{Riess2006a}. This dataset contains 206 supernovae  (including \emph{most but not all} of the {\sf legacy05} supernovae) and reaches out considerably further in redshift, with one outlier at  $z=1.755$,  corresponding to $y=0.6370$. Though the dataset is considerably more extensive it is unfortunately heterogeneous --- combining observations from five different observing platforms over almost a decade. In some cases full data on the operating characteristics of the telescopes used does not appear to be publicly available. The issue of data inhomogeneity has been specifically addressed by Nesseris and Perivolaropoulos in~\cite{heterogeneous}. (For related discussion, see also~\cite{paddy3}.) 
In the {\sf gold06} dataset one is presented with distance moduli and total uncertainty estimates, in particular, including the intrinsic dispersion.

A particular point of interest is that the HST-based high-$z$ supernovae previously published in the gold04 dataset~\cite{gold} have their estimated distances reduced by approximately $5\%$ (corresponding to $\Delta\mu_D = 0.10$), due to a better understanding of nonlinearities in the photodetectors.~\footnote{Changes in stellar magnitude are related to  changes in luminosity distance via equations \eref{E:mu} and \eref{E:d}. Explicitly $\Delta(\ln d_L) = \ln10\; \Delta \mu_D /5$, so that for a given uncertainty in magnitude the  corresponding luminosity distances are multiplied by a factor $10^{\Delta \mu_D/5}$. Then 0.10 magnitudes $\to$ 4.7\% $\approx$ 5\%, and similarly 0.19 magnitudes $\to$ 9.1\%. }
Furthermore, the authors of~\cite{Riess2006a} incorporate (most of) the supernovae in the legacy dataset~\cite{legacy, legacy-url}, but do so in a modified manner by  reducing their estimated distance moduli by  $\Delta\mu_D = 0.19$    (corresponding naively to a $9.1\%$ reduction in luminosity distance) --- however this is only a change in the normalization used in reporting the data, not a physical change in distance. Based on revised modelling of the light curves, and ignoring the question of overall normalization, the overlap between the {\sf gold06} and {\sf legacy05} datasets is argued to be consistent to within $0.5\%$~\cite{Riess2006a}. 

The critical point is this: Since one is still seeing $\approx 5\%$ variations in estimated supernova distances on a two-year timescale, this strongly suggests that the unmodelled systematic uncertainties (the so-called ``unknown unknowns'') are not yet fully under control in even the most recent data. It would be prudent to retain a systematic uncertainty budget of at least $5\%$  (more specifically,  $\Delta\mu_D = 0.10$), and not to place too much credence in any result that is not robust under possible systematic recalibrations of this magnitude. Indeed the authors of~\cite{Riess2006a} state:
\begin{itemize}
\item ``... we adopt a limit on redshift-dependent 
systematics to be 5\% per $\Delta z = 1$'';
\item ``At present, none of the \emph{known}, well-studied sources of systematic error rivals the 
statistical errors presented here.''
\end{itemize}
We shall have more to say about possible systematic uncertainties, both ``known unknowns'' and ``unknown unknowns'' later in this article.

\begin{figure}[htb]
\begin{center}
\includegraphics[width=12cm]{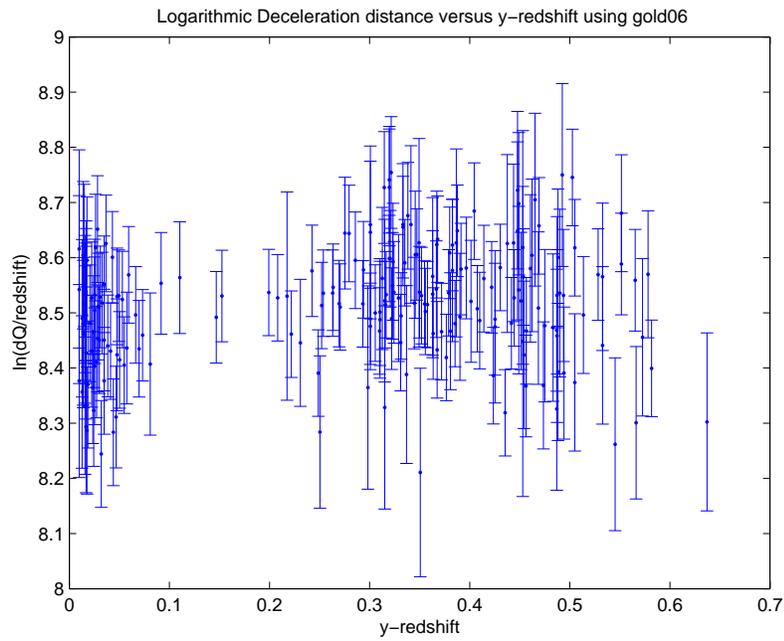}
\end{center}
\caption{\label{F:ln_dQ_gold06}
The normalized logarithm of the deceleration distance, $\ln(d_Q/[y \hbox{ Mpc}])$, as a function of the $y$-redshift using the {\sf gold06} dataset~\cite{gold, Riess2006a}.}
\end{figure}

\begin{figure}[htb]
\begin{center}
\includegraphics[width=12cm]{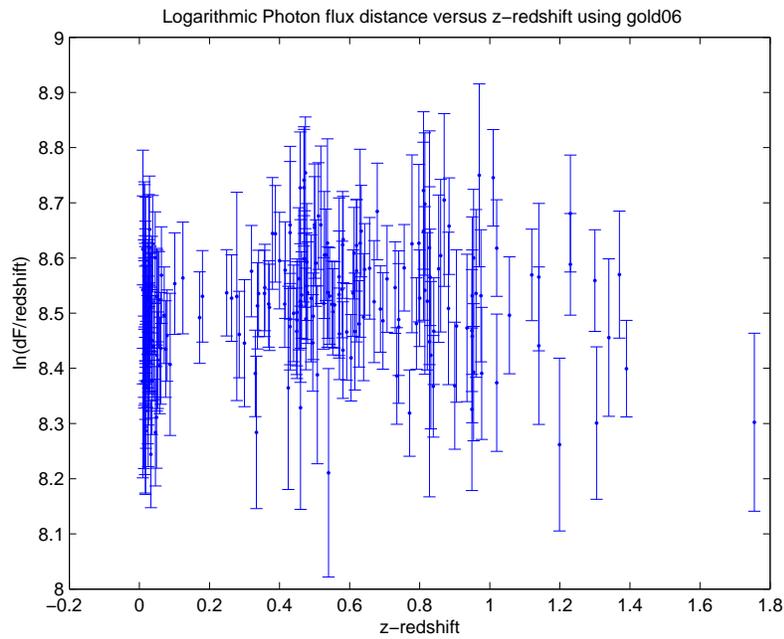}
\end{center}
\caption{\label{F:ln_dF_gold06}
The normalized logarithm of the photon flux distance, $\ln(d_F/[z \hbox{ Mpc}])$, as a function of the $z$-redshift using the {\sf gold06} dataset~\cite{gold, Riess2006a}.}
\end{figure}

To orient oneself, figure  \ref{F:ln_dQ_gold06} again focusses on the normalized logarithm of the deceleration distance $d_Q(y)$ as a function of $y$-redshift. Similarly, figure \ref{F:ln_dF_gold06}  focusses on the normalized logarithm of the photon flux distance $d_F(z)$ as a function of $z$-redshift. 
Visually, these curves are again very close to flat out to $y\approx 0.4$ and  $z\approx 0.4$ respectively, which seems to imply an ``eyeball estimate'' that $q_0\approx 0$. Again, this gives one a feel for just how tricky it is to reliably estimate the deceleration parameter $q_0$ from the data. 

Note the outlier at   $y=0.6370$, that is, $z=1.755$.
In particular,  observe that  adopting the $y$-redshift in place of the $z$-redshift has the effect of pulling this outlier ``closer'' to the main body of data, thus reducing its ``leverage'' effect on any data fitting one undertakes --- apart from the theoretical reasons we have given for preferring the $y$-redshift, (improved convergence behaviour for the Taylor series), the fact that it automatically reduces the leverage of high redshift outliers is a feature that is considered highly desirable purely for statistical reasons. In particular, the method of least-squares is known to be non-robust with respect to outliers. One could implement more robust regression algorithms, but they are not as easy and fast as the classical least-squares method. We have also implemented least-squares regression against a reduced dataset where we have trimmed out the most egregious high-$z$ outlier, and also eliminated the so-called ``Hubble bubble'' for $z < 0.0233$~\cite{bubble1, bubble2}. While the precise numerical values of our estimates for the cosmological parameters then change, there is no great qualitative change to the points we wish to make in this article, nor to the conclusions we will draw.

\subsection{Peculiar velocities}

One point that should be noted for both the {\sf legacy05} and {\sf gold06} datasets is the way that peculiar velocities have been treated. While peculiar velocities would physically seem to be best represented by assigning an uncertainty to the measured redshift, in both these datasets the peculiar velocities have instead been modelled as some particular function of $z$-redshift and then lumped into the reported uncertainties in the distance modulus.  Working with the $y$-redshift \emph{ab initio} might lead one to re-assess the model for the uncertainty due to peculiar velocities. We expect such effects to be small and have not considered them in detail.

\section{Data fitting: Statistical uncertanties}

We shall now compare and contrast the results of multiple least-squares fits to the different notions of cosmological distance, using the two distinct redshift parameterizations discussed above. Specifically, we use a finite-polynomial truncated Taylor series as our model, and perform classical least-squares fits. This is effectively a test of the robustness of the data-fitting procedure, testing it for model dependence. For general background information see~\cite{Bevington, basic1, basic2, basic3, basic4, transformation, nonlinear}.

\subsection{Finite-polynomial truncated-Taylor-series fit}
Working (for purposes of the presentation) in terms of $y$-redshift, the various distance scales can be fitted to finite-length power-series polynomials $d(y)$ of the form
\begin{equation}
P(y): \quad d(y)= \sum_{j=0}^{n} a_j \; y^j,
\end{equation}
where the coefficients $a_j$ all have the dimensions of distance. In contrast, logarithmic fits are of the form
\begin{equation}
P(y):  \quad \ln[d(y)/(y \hbox{ Mpc})]= \sum_{j=0}^{n} b_j \; y^j,
\end{equation}
where the coefficients $b_j$ are now all dimensionless. By fitting to finite polynomials we are implicitly making the assumption that the higher-order coefficients are all exactly zero --- this does then implicitly enforce assumptions regarding the higher-order time derivatives $\d^m a/\d t^m$ for $m>n$, but there is no way to avoid making at least some assumptions of this type~\cite{Bevington, basic1, basic2, basic3, basic4, transformation, nonlinear}.

The method of least squares requires that we minimize
\begin{equation}
\chi^2= \sum_{I=1}^N \left( \frac{P_I-P(y_I)}{\sigma_I} \right)^2,
\end{equation}
where the $N$ data points $(y_I, P_I)$ represent the relevant function $P_I=f(\mu_{D,I},y_I)$ of the distance modulus $\mu_{D,I}$ at corresponding $y$-redshift $y_I$, as inferred from some specific supernovae dataset. Furthermore $P(y_I)$ is the finite polynomial model evaluated at $y_I$. The $\sigma_I$ are the total statistical uncertainty in $P_I$ (including, in particular, intrinsic dispersion).
The location of the minimum value of $\chi^2$ can be determined by setting the derivatives of $\chi^2$ with respect to each of the coefficients $a_j$ or $b_j$ equal to zero.

Note that the theoretical justification for using least squares assumes that the statistical uncertainties are normally distributed Gaussian uncertainties --- and there is no real justification for this assumption in the actual data.  Furthermore if the data is processed by using some nonlinear transformation, then in general Gaussian uncertainties will not remain Gaussian --- and so even if the untransformed uncertainties are Gaussian the theoretical justification for using least squares is again undermined unless the scatter/uncertainties are small, [in the sense that $\sigma \ll f''(x)/f'(x)$], in which case one can appeal to a local linearization of the nonlinear data transformation $f(x)$ to deduce approximately Gaussian uncertainties~\cite{Bevington, basic1, basic2, basic3, basic4, transformation, nonlinear}. As we have already seen, in figures \ref{F:ln_dQ_legacy05}--\ref{F:ln_dF_gold06}, there is again no real justification for this ``small scatter'' assumption in the actual data --- nevertheless, in the absence of any clearly better data-fitting prescription, least squares is the standard way of proceeding. More statistically sophisticated techniques, such as ``robust regression'', have their own distinct draw-backs and, even with weak theoretical underpinning, $\chi^2$ data-fitting is still  typically the technique of choice~\cite{Bevington, basic1, basic2, basic3, basic4, transformation, nonlinear}.

We have performed least squares analyses, both linear in distance and logarithmic in distance, for all of the distance scales discussed above, $d_L$, $d_F$, $d_P$, $d_Q$, and $d_A$, both in terms of $z$-redshift and $y$-redshift,  for finite polynomials from $n=1$ (linear) to $n=7$ (septic). We stopped at $n=7$ since beyond that point the least squares algorithm was found to become numerically unstable due to the need to invert a numerically ill-conditioned matrix --- this ill-conditioning is actually a well-known feature of high-order least-squares polynomial fitting. We carried out the analysis to such high order purely as a diagnostic --- we shall soon see that the ``most reasonable'' fits are actually rather low order $n=2$ quadratic fits.

\subsection{$\chi^2$ goodness of fit}
A convenient measure of the goodness of fit is given by the reduced chi-square:
\begin{equation}
\chi^2_{\nu}= \frac{\chi^2}{\nu},
\end{equation}
where the factor $\nu= N-n-1$ is the number of degrees of freedom left after fitting $N$ data points to the $n+1$ parameters. 
If the fitting function is a good approximation to the parent function, then the value of the reduced chi-square should be approximately unity $\chi^2_{\nu}\approx1$. If the fitting function is not appropriate for describing the data, the value of $\chi^2_{\nu}$ will be greater than $1$. Also, ``too good'' a chi-square fit  ($\chi^2_\nu < 1 $) can come from over-estimating the statistical measurement uncertainties. Again, the theoretical justification for this test relies on the fact that one is assuming, without a strong empirical basis, Gaussian uncertainties~\cite{Bevington, basic1, basic2, basic3, basic4, transformation, nonlinear}.
In all the cases we considered, for polynomials of order $n=2$ and above, we found that  $\chi^2_\nu \approx 1$ for the {\sf legacy05} dataset, and $\chi^2_\nu \approx 0.8 < 1$ for the {\sf gold06} dataset. Linear $n=1$ fits often gave high values for $\chi^2_{\nu}$.  We deduce that:
\begin{itemize}
\item It is desirable to keep at least quadratic $n=2$ terms in all data fits.
\item Caution is required when interpreting the reported statistical uncertainties in the {\sf gold06} dataset.
\end{itemize}
(In particular, note that some of the estimates of the statistical uncertainties reported in {\sf gold06} have themselves been  determined through statistical reasoning --- essentially by adjusting  $\chi^2_{\nu}$ to be ``reasonable''.  The effects of such pre-processing become particularly difficult to untangle when one is dealing with a heterogeneous dataset.)

\subsection{$F$-test of additional terms} 
How many polynomial terms do we need to include  to obtain a good approximation to the parent function? 

The difference between two $\chi^2$ statistics is also distributed as $\chi^2$. In particular, if we fit a set of data with a fitting polynomial of $n-1$ parameters, the resulting value of chi-square associated with the deviations about the regression $\chi^2(n-1)$ has $N-n$ degrees of freedom. If we add another term to the fitting polynomial, the corresponding value of chi-square $\chi^2(n)$ has $N-n-1$ degrees of freedom. The difference between these two follows the $\chi^2$ distribution with one degree of freedom.

The $F_{\chi}$ statistic follows a $F$ distribution with $\nu_1=1$ and $\nu_2=N-n-1$,
\begin{equation}
F_{\chi}=\frac{\chi^2(n-1)-\chi^2(n)}{\chi^2(n)/(N-n-1)}.
\end{equation}
This ratio is a measure of how much the additional term has improved the value of the reduced chi-square.  $F_{\chi}$ should be small when the function with $n$ coefficients does not significantly improve the fit over the polynomial fit with $n-1$ terms.

In all the cases we considered, the $F_\chi$ statistic was not significant when one proceeded beyond $n=2$. 
We deduce that:
\begin{itemize}
\item It is statistically meaningless to go beyond $n=2$ terms in the data fits.
\item This means that one can \emph{at best} hope to estimate the deceleration parameter and the jerk (or more precisely the combination $j_0+\Omega_0$). There is no meaningful hope of estimating the snap parameter from the current data.
\end{itemize}

\subsection{Uncertainties in the coefficients $a_j$ and $b_j$}
From the fit one can determine the standard deviations $\sigma_{a_j}$ and $\sigma_{b_j}$ for the uncertainty of the polynomial coefficients $a_j$ or $b_j$. It is the root sum square of the products of the standard deviation of each data point $\sigma_i$, multiplied by the effect that the data point has on the determination of the coefficient $a_j$~\cite{Bevington}:
\begin{equation}
\sigma_{a_j}^2=\sum_I \left[ \sigma_I^2 \left(  \frac{\partial a_j}{\partial P_I}\right)^2 \right] .
\end{equation}
Similarly the covariance matrix between the estimates of the coefficients in the polynomial fit is
\begin{equation}
\sigma_{a_j a_k}^2=\sum_I \left[ \sigma_I^2 \left(  \frac{\partial a_j}{\partial P_I}\right)   
\left(  \frac{\partial a_k}{\partial P_I}\right)  \right] .
\end{equation}
Practically, the $\sigma_{a_j}$ and covariance matrix $\sigma_{a_j a_k}^2$ 
are determined as follows~\cite{Bevington}:
\begin{itemize}

\item Determine the so-called \emph{curvature matrix} $\alpha$ for our specific polynomial model, where the coefficients are given by
\begin{equation}
\alpha_{jk}=\sum_I \left[ \frac{1}{\sigma_I^2} \; (y_I)^j\; (y_I)^k  \right].
\end{equation}

\item Invert the symmetric matrix $\alpha$ to obtain the so-called \emph{error matrix} $\epsilon$:
\begin{equation}
\epsilon=\alpha^{-1}.
\end{equation}

\item The uncertainty and covariance in the coefficients $a_j$ is characterized by:
\begin{equation}
\sigma_{a_j}^2=\epsilon_{jj}; \qquad \sigma_{a_j a_k}^2 = \epsilon_{jk}.
\end{equation}

\item Finally, for any function $f(a_i)$ of the coefficients $a_i$:
\begin{equation}
\sigma_f = \sqrt{  \sum_{j,k} 
\sigma_{a_j a_k}^2 \; {\partial f\over \partial a_j}  \; {\partial f\over \partial a_k}  }.
\end{equation}

\end{itemize}
Note that these rules for the propagation of uncertainties implicitly assume that the uncertainties are in some suitable sense ``small'' so that a local linearization of the functions $a_j(P_I)$ and $f(a_i)$ is adequate.

Now for each individual element of the curvature matrix
\begin{equation}
0 < {\alpha_{jk}(z)\over(1+z_\mathrm{max})^{2n} } < {\alpha_{jk}(z)\over(1+z_\mathrm{max})^{j+k} } 
< \alpha_{jk}(y) < \alpha_{jk}(z).
\end{equation}
Furthermore the matrices $\alpha_{jk}(z)$ and $\alpha_{jk}(y)$ are both positive definite, and the spectral radius of $\alpha(y)$ is definitely less than the spectral radius of $\alpha(z)$.
After matrix inversion this means that the minimum eigenvalue of the error matrix $\epsilon(y)$ is definitely greater than the minimum eigenvalue of $\epsilon(z)$ --- more generally this tends to make the statistical uncertainties when one works with $y$ greater than the statistical uncertainties when one works with $z$. (However this naive interpretation is perhaps somewhat misleading: It might be more appropriate to say that  the statistical uncertainties when one works with $z$ are anomalously low due to the fact that one has artificially stretched out the domain of the data.)

\subsection{Estimates of the deceleration and jerk}
For all five of the cosmological distance scales discussed in this article, we have calculated the coefficients $b_j$ for the logarithmic distance fits, and their statistical uncertainties, for a polynomial of order $n=2$ in both the $y$-redshift and $z$-redshift, for both the {\sf legacy05} and {\sf gold06} datasets. The constant term $b_0$ is (as usual in this context) a ``nuisance term'' that depends on an overall luminosity calibration that  is not relevant to the questions at hand. These coefficents are then converted to estimates of the deceleration parameter $q_0$ and the combination ($j_0+\Omega_0$)  involving the jerk. A particularly nice feature of the logarithmic distance fits is that logarithmic distances are linearly related to the reported distance modulus. So assumed Gaussian errors in the distance modulus remain Gaussian when reported in terms of logarithmic distance --- which then evades one potential problem source --- whatever is going on in our analysis it is \emph{not} due to the nonlinear transformation of Gaussian errors. We should also mention that for both the {\sf lagacy05} and {\sf gold06} datasets the uncertainties in $z$ have been folded into the reported values of the distance modulus: The reported values of redshift (formally) have no uncertainties associated with them, and so the nonlinear transformation $y\leftrightarrow z$ does not (formally) affect the assumed Gaussian distribution  of the errors.

The results are presented in tables \ref{T:q_y_legacy05}--\ref{T:q_z_gold06}.  Note that even after we have extracted these numerical results there is still a considerable amount of interpretation that has to go into  understanding their physical implications. In particular note that the differences between the various models, (Which distance do we use? Which version of redshift do we use? Which dataset do we use?), often dwarf the statistical uncertainties within any particular model.

\begin{table}[htdp]
\caption{Deceleration and jerk parameters ({\sf legacy05} dataset, $y$-redshift).}
\begin{center}
\begin{tabular}{|c|c|c|}
\hline
distance & $q_0$ & $j_0+\Omega_0$ \\
\hline
$d_L$ & $-0.47\pm 0.38$ & $-0.48\pm3.53$ \\
$d_F$ &  $-0.57\pm0.38$ & $+1.04\pm3.71$ \\
$d_P$&  $-0.66\pm 0.38$& $+2.61\pm3.88$\\
$d_Q$ &  $-0.76\pm0.38$& $+4.22\pm4.04$\\
$d_A$ &  $-0.85\pm0.38 $& $+5.88\pm4.20 $\\
\hline
\end{tabular}
\\[10pt]
{\small With 1-$\sigma$ statistical uncertainties.}
\end{center}
\label{T:q_y_legacy05}
\end{table}%

\begin{table}[htdp]
\caption{Deceleration and jerk parameters ({\sf legacy05} dataset, $z$-redshift).}
\begin{center}
\begin{tabular}{|c|c|c|}
\hline
distance & $q_0$ & $j_0+\Omega_0$ \\
\hline
$d_L$ & $-0.48\pm 0.17$ & $+0.43\pm0.60$ \\
$d_F$ &  $-0.56\pm0.17$ & $+1.16\pm0.65$ \\
$d_P$&  $-0.62\pm 0.17$& $+1.92\pm0.69$\\
$d_Q$ &  $-0.69\pm0.17$& $+2.69\pm0.74$\\
$d_A$ &  $-0.75\pm0.17 $& $+3.49\pm0.79 $\\
\hline
\end{tabular}
\\[10pt]
{\small With 1-$\sigma$ statistical uncertainties.}
\end{center}
\label{T:q_z_legacy05}
\end{table}%

\begin{table}[htdp]
\caption{Deceleration and jerk parameters ({\sf gold06} dataset, $y$-redshift).}
\begin{center}
\begin{tabular}{|c|c|c|}
\hline
distance & $q_0$ & $j_0+\Omega_0$ \\
\hline
$d_L$ & $-0.62\pm 0.29$ & $+1.66\pm2.60$ \\
$d_F$ &  $-0.78\pm0.29$ & $+3.95\pm2.80$ \\
$d_P$&  $-0.94\pm 0.29$& $+6.35\pm3.00$\\
$d_Q$ &  $-1.09\pm0.29$& $+8.87\pm3.20$\\
$d_A$ &  $-1.25\pm0.29 $& $+11.5\pm3.41 $\\
\hline
\end{tabular}
\\[10pt]
{\small With 1-$\sigma$ statistical uncertainties.}
\end{center}
\label{T:q_y_gold06}
\end{table}%

\begin{table}[htdp]
\caption{Deceleration and jerk parameters ({\sf gold06} dataset, $z$-redshift).}
\begin{center}
\begin{tabular}{|c|c|c|}
\hline
distance & $q_0$ & $j_0+\Omega_0$ \\
\hline
$d_L$ & $-0.37\pm 0.11$ & $+0.26\pm0.20$ \\
$d_F$ &  $-0.48\pm0.11$ & $+1.10\pm0.24$ \\
$d_P$&  $-0.58\pm 0.11$& $+1.98\pm0.29$\\
$d_Q$ &  $-0.68\pm0.11$& $+2.92\pm0.37$\\
$d_A$ &  $-0.79\pm0.11 $& $+3.90\pm0.39 $\\
\hline
\end{tabular}
\\[10pt]
{\small With 1-$\sigma$ statistical uncertainties.}
\end{center}
\label{T:q_z_gold06}
\end{table}%

The statistical uncertainties in $q_0$ are independent of the distance scale used because they are linearly related to the statistical uncertainties in the parameter $b_1$, which themselves  depend only on the curvature matrix, which is independent of the distance scale used. In contrast, the statistical uncertainties in ($j_0+\Omega_0$), while they depend linearly the statistical uncertainties in the parameter $b_2$,  depend  nonlinearly on $q_0$ and its statistical uncertainty.

\section{Model-building uncertainties}

The fact that there are such large differences between the cosmological parameters deduced from the different models should give one pause for concern. These differences do not arise from any statistical flaw in the analysis, nor do they in any sense represent any ``systematic'' error, rather they are an intrinsic side-effect of what it means to do a least-squares fit --- to a  finite-polynomial approximate Taylor series --- in a situation where it is physically unclear as to which if any particular measure of ``distance''  is physically preferable, and which particular notion of ``distance'' should be fed into the least-squares algorithm. In \ref{A:ambiguity} we present a brief discussion of the most salient mathematical issues.

The key numerical observations are that the different notions of cosmological distance lead to equally spaced least-squares estimates of the deceleration parameter, with equal statistical uncertainties; the reason for the equal-spacing of these estimates being analytically explainable by the analysis presented in \ref{A:ambiguity}. Furthermore, from the results in \ref{A:ambiguity} we can explicitly calculate the magnitude of this modelling ambiguity as 
\begin{equation}
\left[\Delta q_0\right]_\mathrm{modelling}  = -1 +
 \left[ \sum_I z_I^{i+j} \right]^{-1}_{1j} \; \left[\sum_I z_I^{j} \; \ln(1+z_I)\right] ,
\end{equation}
while the corresponding formula for $y$-redshift is 
\begin{equation}
\left[\Delta q_0\right]_\mathrm{modelling}  = -1 -
 \left[ \sum_I y_I^{i+j} \right]^{-1}_{1j} \; \left[\sum_I y_I^{j} \; \ln(1-y_I)\right] .
\end{equation}
Note that for the quadratic fits we have adopted this requires calculating a $(n+1)\times(n+1)$ matrix, with $\{i,j\} \in \{0,1,2\}$, inverting it, and then taking the inner product between the first row of this inverse matrix and the relevant column vector. The Einstein summation convention is implied on the $j$ index.
For the $z$-redshift (if we were to restrict our $z$-redshift dataset to $z<1$, \eg, using {\sf legacy05} or a truncation of {\sf gold06}) it makes sense to Taylor series expand the logarithm to alternatively yield
\begin{equation}
\left[\Delta q_0\right]_\mathrm{modelling}  =
- \sum_{k= n+1}^\infty {(-1)^{k}\over k}    \left[ \sum_I z_I^{i+j} \right]^{-1}_{1j} \;
\left[ \sum_I z_I^{j+k} \right].
\end{equation}
For the $y$-redshift we do not need this restriction and can simply write
\begin{equation}
\left[\Delta q_0\right]_\mathrm{modelling}  =
 \sum_{k= n+1}^\infty {1\over k}    \left[ \sum_I y_I^{i+j} \right]^{-1}_{1j} \;
\left[ \sum_I y_I^{j+k} \right].
\end{equation}
As an extra consistency check we have independently calculated these quantities (which depend only on the redshifts of the supernovae) and compared them with the spacing we find by comparing  the various least-squares analyses. 
For the $n=2$ quadratic fits these formulae reproduce the spacing reported in tables \ref{T:q_y_legacy05}--\ref{T:q_z_gold06}.
As the order $n$ of the polynomial increases, it was seen that the differences between deceleration parameter estimates based on the different distance measures decreases --- unfortunately the size of the purely statistical uncertainties was simultaneously seen to increase --- this being a side effect of adding terms that are not statistically significant according to the $F$ test.

\emph{Thus to minimize ``model building ambiguities'' one wishes the parameter ``\,$n$'' to be as large as possible, while to minimize statistical uncertainties, one does not want to add statistically meaningless terms to the polynomial.}

Note that if one were to have a clearly preferred physically motivated ``best'' distance this whole model building ambiguity goes away. 
In the absence of a clear physically justifiable preference, the best one can do is to combine the data as per the discussion in \ref{A:combine},  which is based on NIST recommended guidelines~\cite{NIST},
and report an additional model building uncertainty (beyond the traditional purely statistical uncertainty).  

Note that we do limit the modelling uncertainty to that due to considering the five reasonably standard definitions of distance  $d_A$, $d_Q$, $d_P$, $d_F$, and $d_L$. The reasons for this limitation are partially practical (we have to stop somewhere), and partly physics-related (these five definitions of distance have reasonably clear physical interpretations, and there seems to be no good physics reason for constructing yet more notions of cosmological distance).

Turning to the quantity ($j_0+\Omega_0$), the different notions of distance no longer yield equally spaced  estimates, nor are the statistical uncertainties equal. This is due to the fact that there is a nonlinear quadratic term involving $q_0$ present in the relation used to convert the polynomial coefficient $b_2$ into the more physical parameter ($j_0+\Omega_0$).  Note that while for each specific model (choice of distance scale and redshift variable) the $F$-test indicates that keeping the quadratic term is statistically significant, the variation among the models is so great as to make measurements of  ($j_0+\Omega_0$) almost meaningless.  The combined results are reported in tables \ref{T:q-stat-model}--\ref{T:j-stat-model}. Note that these tables do not yet include \emph{any} budget for ``systematic'' uncertainties. 

\begin{table}[htdp]
\caption{Deceleration parameter summary: Statistical plus modelling.}
\begin{center}
\begin{tabular}{|c|c|c|}
\hline
dataset & redshift & 
$q_0\pm\sigma_\mathrm{statistical}\pm\sigma_\mathrm{modelling}$   \\
\hline
{\sf legacy05} & $y$ & $-0.66\pm0.38\pm0.13$  \\
{\sf legacy05} & $z$ & $-0.62\pm0.17\pm0.10$ \\
{\sf gold06}     & $y$ & $-0.94\pm0.29\pm0.22$  \\
{\sf gold06}     & $z$ & $-0.58\pm0.11\pm0.15$\\
\hline
\end{tabular}
\\[10pt]
{\small With 1-$\sigma$ statistical uncertainties and 1-$\sigma$ model building uncertainties,\\
no budget for ``systematic'' uncertainties.}
\end{center}
\label{T:q-stat-model}
\end{table}%

\begin{table}[htdp]
\caption{Jerk parameter summary: Statistical plus modelling.}
\begin{center}
\begin{tabular}{|c|c|c|}
\hline
dataset & redshift & 
$(j_0+\Omega_0)\pm\sigma_\mathrm{statistical}\pm\sigma_\mathrm{modelling}$  \\
\hline
{\sf legacy05} & $y$ & $+2.65\pm3.88\pm2.25$  \\
{\sf legacy05} & $z$ & $+1.94\pm0.70\pm1.08$ \\
{\sf gold06 }    & $y$ & $+6.47\pm3.02\pm3.48$  \\
{\sf gold06 }    & $z$ & $+2.03\pm0.31\pm1.29$\\
\hline
\end{tabular}
\\[10pt]
{\small With 1-$\sigma$ statistical uncertainties and 1-$\sigma$ model building uncertanties,\\
no budget for ``systematic'' uncertainties.}
\end{center}
\label{T:j-stat-model}
\end{table}%

Again, we reiterate the fact that there are distressingly large differences between the cosmological parameters deduced from the different models --- this should give one pause for concern above and beyond the purely formal statistical uncertainties reported herein.

\section{Systematic uncertainties}

Beyond the statistical uncertainties and model-building uncertainties we have so far considered lies the issue of systematic uncertainties. Systematic uncertainties are  extremely difficult to quantify in cosmology, at least when it comes to distance measurements --- see for instance the relevant discussion in~\cite{Riess2006a, Riess2006b}, or in~\cite{essence}.
What is less difficult to quantify, but still somewhat tricky, is the extent to which systematics propagate through the calculation. 

\subsection{Major philosophies underlying the analysis of statistical uncertainty}
\label{SS:philosophies}

When it comes to dealing with systematic uncertainties there are two major philosophies on how to report and analyze them:
\begin{itemize}

\item Treat all systematic uncertainties \emph{as though} they were purely statistical and report 1-sigma ``effective standard uncertainties''. In propagating systematic uncertainties treat them \emph{as though} they were purely statistical and uncorrelated with the usual statistical uncertainties. In particular, this implies that one is to add estimated systematic and statistical uncertainties in quadrature
\begin{equation}
\sigma^2_\mathrm{combined} = \sqrt{ \sigma_\mathrm{statistical}^2 + \sigma_\mathrm{systematic}^2 }.
\end{equation}
This manner of treating the systematic uncertainties is that currently recommended by NIST~\cite{NIST},
this recommendation itself being based on ISO, CPIM, and BIPM recommendations.  This is also the language most widely used within the supernova community, and in particular in discussing the {\sf gold05} and {\sf legacy05} datasets~\cite{legacy, legacy-url, gold, Riess2006a, Riess2006b}, so we shall standardize our language to follow these norms.

\item An alternative manner of dealing with systematics (now deprecated) is to carefully segregate systematic and statistical effects, somehow estimate ``credible bounds'' on the systematic uncertainties, and then propagate the systematics through the calculation --- if necessary using \emph{interval arithmetic} to place ``credible bounds'' on the final reported systematic uncertainty. The measurements results would then be reported as a number with two independent sources of uncertainty --- the statistical and systematic uncertainties, and within this philosophy there is no justification for adding statistical and systematic effects in quadrature. 

\end{itemize}
It is important to realise that the systematic uncertainties reported in  {\sf gold05} and {\sf legacy05}  are of the first type: effective equivalent 1-sigma error bars~\cite{legacy, legacy-url, gold, Riess2006a, Riess2006b}. These reported uncertainties are based on what in the supernova community are referred to as ``known unknowns''.

(The NIST guidelines~\cite{NIST} also recommend that all uncertainties estimated by statistical methods should be denoted by the symbol $s$, not $\sigma$, and that uncertainties estimated by non-statistical methods, and combined overall uncertainties, should be denoted by the symbol $u$ --- but this is rarely done in practice, and we shall follow the traditional abuse of notation and continue to use $\sigma$ throughout.)

\subsection{Deceleration}
For instance, assume we can measure distance moduli to within a systematic uncertainty $\Delta \mu_\mathrm{systematic}$ over a redshift range $\Delta (\mathrm{redshift})$.  If all the measurements are biased high, or all are biased low, then the systematic uncertainty would affect the Hubble parameter $H_0$, but would not in any way disturb the deceleration parameter $q_0$.  However there may be a systematic drift in the bias as one scans across the range of observed redshifts. 
The worst that could plausibly happen is that all measurements are systematically biased high at one end of the range, and biased low at the other end of the range. For data collected over a finite width $\Delta(\mathrm{redshift})$,  this ``worst plausible'' situation leads to a systematic uncertainty in the slope of
\begin{equation}
\label{E:sigma_systematic}
\Delta\left[{d\mu\over dz} \right]_\mathrm{systematic} = 
{2 \; \Delta\mu_\mathrm{systematic}\over\Delta (\mathrm{redshift})},
\end{equation}
which then propagates to an uncertainty in the deceleration parameter of
\begin{equation}
\fl
\sigma_\mathrm{systematic} = 
{2 \ln10\over5} \;\Delta\left[{d\mu\over dz} \right]_\mathrm{systematic} =
{4 \ln10\over 5} \; { \Delta\mu_\mathrm{systematic}\over\Delta (\hbox{redshift})}
\approx
1.8  \;\; { \Delta\mu_\mathrm{systematic}\over\Delta (\hbox{redshift})}.
\end{equation}
For the situation we are interested in, if we take at face value the reliability of the assertion ``...we adopt a limit on redshift-dependent 
systematics to be 5\% per $\Delta z = 1$''~\cite{Riess2006a}, meaning up to 2.5\% high at one end of the range and up to 2.5\% low at the other end of the range. A 2.5\% variation in distance then corresponds,
via  $\Delta\mu_D= 5 \Delta(\ln d_L)/\ln10$, to an uncertainty $\Delta\mu_\mathrm{systematic} = 0.05$ in stellar magnitude.   So, (taking $\Delta z = 1$), one has to face  the somewhat sobering estimate that the ``equivalent 1-$\sigma$ uncertainty'' for the deceleration parameter $q_0$ is
\begin{equation}
\sigma_\mathrm{systematic} = 0.09.
\end{equation}
When working with $y$-redshift, one really should reanalyze the entire corpus of data from first principles --- failing that, (not enough of the raw data is publicly available), we shall simply observe that
\begin{equation}
{\d z\over\d y} \to 1 \qquad \hbox{as} \qquad y \to 0,
\end{equation}
and use this as a justification for assuming that the systematic uncertainty in $q_0$ when using $y$-redshift is the same as when using $z$-redshift.

\subsection{Jerk}
Turning to systematic uncertainties in the jerk,  the worst that could plausibly happen is that all measurements are systematically biased high at both ends of the range, and biased low at the middle, (or low at both ends and high in the middle), leading to a systematic uncertainty in the second derivative  of
\begin{equation}
\label{E:jerk}
{1\over 2} \; \Delta\left[{d^2\mu\over dz^2} \right]_\mathrm{systematic} \; \left[{\Delta (\hbox{redshift})}\over2\right]^2 
=  2 \Delta\mu_\mathrm{systematic},
\end{equation}
where we have taken the second-order term in the Taylor expansion around the midpoint of the redshift range, and asked that it saturate the estimated systematic error $2 \Delta \mu_\mathrm{systematic}$.
This implies
\begin{equation}
\Delta\left[{d^2\mu\over dz^2} \right]_\mathrm{systematic} = {16 \; \Delta\mu_\mathrm{systematic}\over\Delta (\hbox{redshift})^2},
\end{equation}
which then propagates to an uncertainty in the jerk parameter ($j_0+\Omega_0$) of \emph{at least}
\begin{equation}
\fl 
\sigma_{\mathrm{systematic}} \geq 
{3\ln10\over 5} \; \Delta\left[{d^2\mu\over dz^2} \right]_\mathrm{systematic}
=
{48 \ln10\over 5} \; { \Delta\mu_\mathrm{systematic}\over\Delta (\hbox{redshift})^2 }
\approx
22  \; { \Delta\mu_\mathrm{systematic}\over\Delta (\hbox{redshift})^2}.
\end{equation}
There are additional contributions to the systematic uncertainty arising from terms linear and quadratic in $q_0$. They do not seem to be important in the situations we are interested in so we content ourselves with the single term estimated above. Using $\Delta\mu_\mathrm{systematic} = 0.05$ and $\Delta z=1$  we see that the ``equivalent 1-$\sigma$ uncertainty'' for the combination $ (j_0+\Omega_0)$ is:
\begin{equation}
\sigma_{\mathrm{systematic}} = 1.11.
\end{equation}
Thus direct cosmographic measurements of the jerk parameter are plagued by \emph{very} high systematic uncertainties.
Note that the systematic uncertainties calculated in this section are completely equivalent to those reported in~\cite{Riess2006a}.

\section{Historical estimates of systematic uncertainty}

We  now turn to the question of possible additional contributions to the uncertainty, based on  what the NIST recommendations call ``type B evaluations of uncertainty" --- namely ``any method of evaluation of uncertainty by means other than the statistical analysis of a series of observations''~\cite{NIST}. (This includes effects that  in the supernova community are referred to as ``unknown unknowns", which are \emph{not} reported in any of their estimates of systematic uncertainty.)

The key point here is this: ``A type B evaluation of standard uncertainty is usually based on scientific judgment using all of the relevant information available, which may include: previous measurement data, \emph{etc...}"~\cite{NIST}.  It is this recommendation that underlies what we might wish to call the ``historical" estimates of systematic uncertainty --- roughly speaking, we suggest that in the systematic uncertainty budget it is prudent to keep an extra ``historical uncertainty'' at least as large as the most recent major re-calibration of whatever measurement method you are currently using.

Now this ``historical uncertainty'' contribution to the systematic uncertainty budget that we are advocating is based on 100 years of unanticipated systematic errors (``unknown unknowns'') in astrophysical distance scales --- from Hubble's reliance on mis-calibrated Cephid variables (leading to distance estimates that were about 666\% too large), to last decade's debates on the size of our own galaxy (with up to 15\% disagreements being common), to last year's 5\% shift in the high-$z$ supernova distances~\cite{Riess2006a, Riess2006b} --- and various other re-calibration events in between.  That is, 5\% variations in estimates of cosmological distances  on a 2 year time scale seem common, 10\% on a 10 year time scale, and 500\% or more on an 80 year timescale? A disinterested outside observer does detect a certain pattern here.  (These re-calibrations are of course not all related to supernova measurements, but they are historical evidence of how difficult it is to make reliable distance measurements in cosmology.) Based on the historical evidence we feel that it is currently prudent to budget an additional ``historical uncertainty'' of approximately 5\%  in the distances to the furthest supernovae, (corresponding to 0.10 stellar magnitudes), while for the nearby supernovae we generously budget  a ``historical uncertainty'' of 0\%, based on the fact that these distances have not changed in the last 2 years~\cite{Riess2006a, Riess2006b}.\footnote{Some researchers have argued that the present ``historical'' estimates of uncertainty confuse the notion of ``error'' with that of ``uncertainty''. We disagree. What we are doing here is to use the most recently detected (significant) error to estimate one component of the uncertainty --- this is simply a ``scientific judgment using all of the relevant information available''. We should add that other researchers have argued that our historical uncertainties should be even larger. By using the most recent major re-calibration as our basis for historical uncertainty we feel we are steering a middle course between placing too much \emph{versus} to little credence in the observational data.}

\subsection{Deceleration}

This implies
\begin{equation}
\Delta\left[{d\mu\over dz} \right]_\mathrm{historical} = 
{\Delta\mu_\mathrm{historical}\over\Delta (\mathrm{redshift})}.
\end{equation}
Note the \emph{absence} of a factor 2 compared to equation (\ref{E:sigma_systematic}), this is because in this ``historical'' discussion we have taken the nearby supernovae to be accurately calibrated, whereas in the discussion of systematic uncertainties in equation (\ref{E:sigma_systematic}) both nearby and distant supernovae are subject to ``known unknown'' systematics.
This then propagates to an uncertainty in the deceleration parameter of
\begin{equation}
\fl
\sigma_\mathrm{historical} = 
{2 \ln10\over5} \;\Delta\left[{d\mu\over dz} \right]_\mathrm{historical} =
{2 \ln10\over 5} \; { \Delta\mu_\mathrm{historical}\over\Delta (\hbox{redshift})}
\approx
0.9  \;\; { \Delta\mu_\mathrm{historical}\over\Delta (\hbox{redshift})}.
\end{equation}
Noting that a 5\% shift in luminosity distance is equivalent to an uncertainty of $\Delta\mu_\mathrm{historical} = 0.10$ in stellar magnitude, this implies an
 ``equivalent 1-$\sigma$ uncertainty'' for the deceleration parameter $q_0$ is
\begin{equation}
\sigma_\mathrm{historical} = 0.09.
\end{equation}
This (coincidentally) is equal to the systematic uncertainties based on ``known unknowns''. 

\subsection{Jerk}

Turning to the second derivative a similar analysis implies
\begin{equation}
{1\over 2} \; \Delta\left[{d^2\mu\over dz^2} \right]_\mathrm{historical} \; \Delta (\hbox{redshift})^2 
=  \Delta\mu_\mathrm{historical}.
\end{equation}
Note the absence of various factors of 2 as compared to equation \eref{E:jerk}. This is because we are now assuming that for ``historical'' purposes the nearby supernovae are accurately calibrated and it is only the distant supernovae that are potentially uncertain --- thus in estimating the historical uncertainty the second-order term in the Taylor series is now to be saturated using the entire redshift range.
Thus
\begin{equation}
\Delta\left[{d^2\mu\over dz^2} \right]_\mathrm{historical} = {2 \; \Delta\mu_\mathrm{historical}\over\Delta (\hbox{redshift})^2},
\end{equation}
which then propagates to an uncertainty in the jerk parameter of \emph{at least}
\begin{equation}
\fl
\sigma_{\mathrm{historical}} \geq 
{3\ln10\over 5} \; \Delta\left[{d^2\mu\over dz^2} \right]_\mathrm{historical}
=
{6 \ln10\over 5} \; { \Delta\mu_\mathrm{historical}\over\Delta (\hbox{redshift})^2 }
\approx
2.75  \; { \Delta\mu_\mathrm{historical}\over\Delta (\hbox{redshift})^2}.
\end{equation}
Again taking $\Delta\mu_\mathrm{historical} = 0.10$ this implies an
 ``equivalent 1-$\sigma$ uncertainty'' for the  combination $j_0+\Omega_0$ is
\begin{equation}
\sigma_\mathrm{historical} = 0.28.
\end{equation}
Note that this is (coincidentally) one quarter the size of the systematic uncertainties based on ``known unknowns'', and is still quite sizable.

The systematic and historical uncertainties are now reported in tables \ref{T:q-all}--\ref{T:j-all}. The estimate for systematic uncertainties are equivalent to those presented in~\cite{Riess2006a}, which is largely in accord with related sources~\cite{legacy, legacy-url, gold}. Our estimate for ``historical'' uncertainties is likely to be more controversial --- with, we suspect, many cosmologists arguing that our estimates are too generous --- and that $\sigma_\mathrm{historical}$ should perhaps be \emph{even larger} than we have estimated.  What is not (or should not) be controversial is the need for \emph{some} estimate of $\sigma_\mathrm{historical}$. Previous history should not be ignored, and as the NIST guidelines emphasize, previous history is an essential and integral part of making the scientific judgment as to what the overall uncertainties are.

\begin{table}[htdp]
\caption{Deceleration parameter summary: \\ Statistical, modelling, systematic, and historical.}
\begin{center}
\begin{tabular}{|c|c|c|}
\hline
dataset & redshift & 
$q_0\pm\sigma_\mathrm{statistical}\pm\sigma_\mathrm{modelling}
\pm\sigma_\mathrm{systematic}\pm\sigma_\mathrm{historical}$   \\
\hline
{\sf legacy05} & $y$ & $-0.66\pm0.38\pm0.13\pm0.09\pm0.09$  \\
{\sf legacy05} & $z$ & $-0.62\pm0.17\pm0.10\pm0.09\pm0.09$ \\
{\sf gold06}     & $y$ & $-0.94\pm0.29\pm0.22\pm0.09\pm0.09$  \\
{\sf gold06}     & $z$ & $-0.58\pm0.11\pm0.15\pm0.09\pm0.09$\\
\hline
\end{tabular}
\\[10pt]
{\small With 1-$\sigma$ effective statistical uncertainties for all components.}
\end{center}
\label{T:q-all}
\end{table}%

\begin{table}[htdp]
\caption{Jerk parameter summary: \\
Statistical, modelling, systematic, and historical.}
\begin{center}
\begin{tabular}{|c|c|c|}
\hline
dataset & redshift & 
$(j_0+\Omega_0)\pm\sigma_\mathrm{statistical}\pm\sigma_\mathrm{modelling} 
\pm\sigma_\mathrm{systematic}\pm\sigma_\mathrm{historical}$  \\
\hline
{\sf legacy05} & $y$ & $+2.65\pm3.88\pm2.25\pm 1.11 \pm 0.28$  \\
{\sf legacy05} & $z$ & $+1.94\pm0.70\pm1.08\pm 1.11 \pm 0.28$ \\
{\sf gold06 }    & $y$ & $+6.47\pm3.02\pm3.48\pm 1.11 \pm 0.28$  \\
{\sf gold06 }    & $z$ & $+2.03\pm0.31\pm1.29\pm 1.11 \pm 0.28$\\
\hline
\end{tabular}
\\[10pt]
{\small With 1-$\sigma$ effective statistical uncertainties for all components.}
\end{center}
\label{T:j-all}
\end{table}%

\section{Combined uncertainties}
We now combine these various uncertainties, purely statistical, modelling, ``known unknown'' systematics, and ``historical'' (``unknown unknowns''). Adopting the NIST philosophy of dealing with systematics, these uncertainties are to be added in quadrature~\cite{NIST}. Including all 4 sources of uncertainty we have discussed:
\begin{equation}
\sigma_\mathrm{combined} = \sqrt{ \sigma_\mathrm{statistical}^2 + \sigma_\mathrm{modelling}^2 
 + \sigma_\mathrm{systematic}^2  + \sigma_\mathrm{historical}^2}.
\end{equation}
That the statistical and modelling uncertainties should be added in quadrature is clear from their definition.  Whether or not systematic and historical uncertainties should be treated this way is very far from clear, and implicitly presupposes that there are no correlations between the systematics and the statistical uncertainties --- within the ``credible bounds'' philosophy for estimating systematic uncertainties there is no justification for such a step. Within the ``all errors are effectively statistical'' philosophy adding in quadrature is standard and in fact recommended --- this is what is done in current supernova analyses, and we shall continue to do so here. The combined uncertainties $\sigma_\mathrm{combined}$ are reported in tables \ref{T:q-combined}--\ref{T:j-combined}.

\section{Expanded uncertainty}

An important concept under the NIST guidelines is that of   ``expanded uncertainty''
\begin{equation}
U_k = k \; \sigma_\mathrm{combined}.
\end{equation}
Expanded uncertainty is used when for either scientific or legal/regulatory reasons one wishes to be ``certain'' that the actual physical value of the quantity being measured lies within the stated range. We shall take $k=3$,  this being equivalent to the well-known particle physics aphorism ``if it's not three-sigma, it's not physics''.  Note that this is not an invitation to randomly multiply uncertainties by 3, rather it is a scientific judgment that if one wishes to be $99.5\%$ certain that something is or is not happening one should look for a 3-sigma effect. Bitter experience within the particle physics community has led to the consensus that 3-sigma is the minimum standard one should look for when claiming ``new physics''.\footnote{There is now a growing consensus in the particle physics community that 5-sigma should be the new standard for claiming ``new physics''~\cite{Signal}.} Thus we take
\begin{equation}
U_3 = 3 \; \sigma_\mathrm{combined}.
\end{equation}
The best estimates, combined uncertainties $\sigma_\mathrm{combined}$, and expanded uncertainties $U$, are reported in tables \ref{T:q-combined}--\ref{T:j-combined}.

\begin{table}[htdp]
\caption{Deceleration parameter summary: \\ 
Combined and expanded uncertainties.}
\begin{center}
\begin{tabular}{|c|c|c|c|}
\hline
dataset & redshift & 
$q_0\pm\sigma_\mathrm{combined} $ & $q_0 \pm U_3$   \\
\hline
{\sf legacy05} & $y$ & $-0.66\pm0.42$  &$-0.66\pm1.26$\\
{\sf legacy05} & $z$ & $-0.62\pm0.23$ & $-0.62\pm0.70$ \\
{\sf gold06}     & $y$ & $-0.94\pm0.39$  &$-0.94\pm1.16$ \\
{\sf gold06}     & $z$ & $-0.58\pm0.23$ & $-0.58\pm0.68$ \\
\hline
\end{tabular}
\end{center}
\label{T:q-combined}
\end{table}%

\begin{table}[htdp]
\caption{Jerk parameter summary: \\
Combined and expanded uncertainties.}
\begin{center}
\begin{tabular}{|c|c|c|c|}
\hline
dataset & redshift & 
$(j_0+\Omega_0)\pm\sigma_\mathrm{combined}$ & $(j_0+\Omega_0)\pm U_3$  \\
\hline
{\sf legacy05} & $y$ & $+2.65\pm4.63$ & $+2.65\pm13.9$\\
{\sf legacy05} & $z$ & $+1.94\pm1.72$ & $+1.94\pm5.17$\\
{\sf gold06 }    & $y$ & $+6.47\pm4.75$ &  $+6.47\pm14.2$\\
{\sf gold06 }    & $z$ & $+2.03\pm1.75$ &  $+2.03\pm5.26$\\
\hline
\end{tabular}
\end{center}
\label{T:j-combined}
\end{table}%

\section{Results}

What can we conclude from this? While the ``preponderance of evidence'' is certainly that the universe is currently accelerating, $q_0<0$, this is not yet a ``gold plated'' result. We emphasise the fact that (as is or should be well known) there is an enormous difference between the two statements:
\begin{itemize}
\item ``the most likely value for the deceleration parameter is negative'', and
\item ``there is significant evidence that the deceleration parameter is negative''.
\end{itemize}
When it comes to assessing whether or not the evidence for an accelerating universe is physically significant, the first rule of thumb for combined uncertainties is the well known aphorism ``if it's not three-sigma, it's not physics''. The second rule is to be conservative in your systematic uncertainty budget. 
We cannot in good faith conclude that the expansion of the universe is accelerating. It is more likely that the expansion of the universe is accelerating, than that the expansion of the universe is decelerating --- but this is a very long way from having definite evidence in favour of acceleration.
The summary regarding the jerk parameter, or more precisely $(j_0+\Omega_0)$,  is rather grim reading, and indicates the need for considerable caution in interpreting the supernova data.
Note that while use of the $y$-redshift may improve the theoretical convergence properties of the Taylor series, and will not affect the uncertainties in the distance modulus or the various distance measures, it does seem to have an unfortunate side-effect of magnifying statistical uncertainties for the cosmological parameters. 

As previously mentioned, we have further checked the robustness of our analysis  by first excluding the outlier at $z=1.755$, then excluding the so-called ``Hubble bubble'' at $z<0.0233$~\cite{bubble1,bubble2}, and then excluding both --- the precise numerical estimates for the cosmological parameters certainly  change, but the qualitative picture remains as we have painted it here.

\section{Conclusions}

Why do our conclusions seem to be so much at variance with currently perceived wisdom concerning the acceleration of the universe? The main reasons are twofold:
\begin{itemize}

\item Instead of simply picking a single model and fitting the data to it, we have tested the overall robustness of the scenario by encoding the same physics ($H_0$, $q_0$, $j_0$) in multiple different ways ($d_L$, $d_F$, $d_P$, $d_Q$, $d_A$;  using both $z$ and $y$) to test the robustness of the data fitting procedures.

\item We have been much more explicit, and conservative, about the role of systematic uncertainties, and their effects on estimates of the cosmological parameters.

\end{itemize}
If we \emph{only} use the statistical uncertainties and the ``known unknowns'' added in quadrature, then the case for cosmological acceleration is much improved, and is (in some cases we study) ``statistically significant at three-sigma'', but this does not mean that such a conclusion is either robust or reliable. (By ``cherry picking'' the data, and the particular way one analyzes the data, one can find statistical support for almost any conclusion one wants.)  

The modelling uncertainties we have encountered depend on the distance variable one chooses to do the least squares fit ($d_L$, $d_F$, $d_P$, $d_Q$, $d_A$). There is no good physics reason for preferring any one of these distance variables over the others. One can always minimize the modelling uncertainties by going to a higher-order polynomial --- unfortunately at the price of unacceptably increasing the statistical uncertainties --- and we have checked that this makes the overall situation worse.  This does however suggest that things might improve if the data had smaller scatter and smaller statistical uncertainties: We could then hope that the $F$-test would allow us to go to a cubic polynomial, in which case the dependence on which notion of distance we use for least-squares fitting should decrease.  

\begin{quote}
\emph{We wish to emphasize the point that, regardless of one's views on how to combine formal estimates of uncertainty, the very fact that different distance scales yield data-fits with such widely discrepant values strongly suggests the need for extreme caution in interpreting the supernova data.}
\end{quote}

Though we have chosen to work on a cosmographic framework, and so minimize the number of physics assumptions that go into the model, we expect that similar modelling uncertainties will also plague other more traditional approaches. (For instance, in the present-day consensus scenario there is considerable debate as to just when the universe switches from deceleration to acceleration, with different models making different statistical predictions~\cite{Jonsson}.)  One lesson to take from the current analysis is that purely statistical estimates of error, while they can be used to make statistical deductions within the context of a specific model,  are often a bad guide as to the extent to which two different models for the same physics will yield differing estimates for the same physical quantity.

There are a number of other more sophisticated statistical methods that might be applied to the data to possibly improve the statistical situation. For instance, ridge regression, robust regression, and the use of orthogonal polynomials and loess curves. However one should always keep in mind the difference between \emph{accuracy} and \emph{precision}~\cite{Bevington}. More sophisticated statistical analyses may permit one to improve the precision of the analysis, but unless one can further constrain the systematic uncertainties such precise results will be no more accurate than the current  situation. Excessive refinement in the statistical analysis, in the absence of improved bounds on the systematic uncertainties, is counterproductive and grossly misleading. 

However, we are certainly not claiming that all is grim on the cosmological front --- and do not wish our views to be misinterpreted in this regard --- there are clearly parts of cosmology where there is plenty of high-quality data, and more coming in, constraining and helping refine our models.  But regarding some specific cosmological questions the catch cry should still be ``Precision cosmology? Not just yet"~\cite{precision}.

In particular, in order for the current technique to become a tool for precision cosmology, we would need more data, smaller scatter in the data, and smaller uncertainties.  For instance, by performing the $F$-test we found that it was almost always statistically meaningless to go beyond quadratic fits to the data. If one can obtain an improved dataset of sufficient quality for cubic fits to be meaningful, then ambiguities in the deceleration parameter are greatly suppressed.
 
In closing, we strongly encourage readers to carefully contemplate figures \ref{F:ln_dQ_legacy05}--\ref{F:ln_dF_gold06} as an inoculation against over-interpretation of the supernova data.  In those figures we have split off the linear part of the Hubble law (which is encoded in the intercept) and chosen distance variables so that the slope (at redshift zero) of whatever curve one fits to those plots is directly proportional to the acceleration of the universe (in fact the slope is equal to $-q_0/2$). Remember that these plots only exhibit the statistical uncertainties. Remembering that we prefer to work with natural logarithms, not stellar magnitudes, one should add systematic uncertainties of $\pm [\ln(10)/5]\times(0.05)\approx 0.023$ to these statistical error bars, presumably in quadrature. 
Furthermore a good case can be made for adding an additional ``historical'' uncertainty, using the past history of the field to estimate the ``unknown unknowns''.

\begin{quote}
\emph{Ultimately however, it is the fact that  figures \ref{F:ln_dQ_legacy05}--\ref{F:ln_dF_gold06}  do \emph{not} exhibit any overwhelmingly obvious trend that makes it so difficult to make a robust and reliable estimate of the sign of the deceleration parameter.}
\end{quote}

\appendix
\section[\\ Some ambiguities in least-squares fitting ]{Some ambiguities in least-squares fitting}
\label{A:ambiguity}
Let us suppose we have a function $f(x)$, and want to estimate $f(x)$ and its derivatives at zero via least squares. 
For any $g(x)$ we have a mathematical identity
\begin{equation}
f(x) =  [f(x) - g(x)] + g(x),
\end{equation}
and for the derivatives
\begin{equation}
f^{(m)}(0)  = [f-g]^{(m)}(0) + g^{(m)}(0).
\end{equation}
Adding and subtracting the same function $g(x)$ makes no difference to the underlying function $f(x)$, but it may modify the least squares estimate for that function. That is: Adding and subtracting a \emph{known} function to the data \emph{does not commute} with the process of performing a finite-polynomial least-squares fit.
Indeed, let us approximate
\begin{equation}
[f(x) - g(x)] = \sum_{i=0}^n b_{f-g,i} \; x^i +  \epsilon.
\end{equation}
Then given a set of observations at points $(f_I,x_I)$ we have (in the usual manner) the equations (for simplicity of the presentation all statistical uncertainties $\sigma$ are set equal for now) 
\begin{equation}
[f_I - g(x_I)] = \sum_{i=0}^n \hat b_{f-g,i} \; x_I^i + \epsilon_I,
\end{equation}
where we want to minimize
\begin{equation}
\sum_I  |\epsilon_I |^2.
\end{equation}
This leads to
\begin{equation}
\sum_I [f_I - g(x_I)]  x_I^j = \sum_{i=0}^n \hat b_{f-g,i} \; \sum_I x_I^{i+j},
\end{equation}
whence
\begin{equation}
\hat b_{f-g,i} = \left[ \sum_I x_I^{i+j} \right]^{-1} \; \sum_I [f_I - g(x_I)]  x_I^j,
\end{equation}
where the square brackets now indicate an $(n+1)\times(n+1)$ matrix, and there is an implicit sum on the $j$ index as per the Einstein summation convention.
But we can re-write this as 
\begin{equation}
\hat b_{f-g,i} =    \hat b_{f,i}  - \left[ \sum_I x_I^{i+j} \right]^{-1} \; \sum_I [g(x_I)]  x_I^j,
\end{equation}
relating the least-squares estimates of $b_{f,i}$ and $b_{f-g,i}$.  Note that by construction $i \leq n$.
If we now use this to estimate $f^{(i)}(0)$, we see:
\begin{equation}
\label{E:A9}
\hat f^{(i)}_{[f-g]+g}(0) = \hat f^{(i)}_{f-g}(0) + g^{(i)}(0),
\end{equation}
whence
\begin{equation}
\label{E:A10}
\hat f^{(i)}_{[f-g]+g}(0) = \hat f^{(i)}(0) -  i! \; \left[ \sum_I x_I^{i+j} \right]^{-1} \; \sum_I [g(x_I)]  x_I^j  + g^{(i)}(0),
\end{equation}
where $\hat f^{(i)}(0)$ is the ``naive" estimate of $f^{(i)}(0)$ obtained by simply fitting a polynomial to $f$ itself, and $ \hat f^{(i)}_{[f-g]+g}(0) $ is the ``improved'' estimate obtained by first subtracting $g(x)$, fitting $f(x)-g(x)$ to a polynomial, and then adding $g(x)$ back again.  Note the formula for the shift of the estimate of the $i$th derivative of $f(x)$ is linear in the function $g(x)$ and its derivatives. In general this is the most precise statement we can make --- the process of finding a truncated Taylor series simply does not commute with the process of performing a least squares fit.

We can gain some additional insight if we  use Taylor's theorem to write
\begin{equation}
g(x) = \sum_{k=0}^\infty {g^{(k)}(0)\over k!} x^k = \sum_{k=0}^n {g^{(k)}(0)\over k!} x^k
+ \sum_{k=n+1}^\infty {g^{(k)}(0)\over k!} x^k,
\end{equation}
where we temporarily suspend concerns regarding convergence of the Taylor series.
Then
\begin{eqnarray}
\fl
\hat f^{(i)}_{[f-g]+g}(0) &=& \hat f^{(i)}(0) + g^{(i)}(0) 
\\
\fl
&& 
-  i! \; \left[ \sum_I x_I^{i+j} \right]^{-1} \; 
\sum_I \left\{   \sum_{k=0}^n {g^{(k)}(0)\over j!} x_I^k + 
\sum_{k= n+1}^\infty {g^{(k)}(0)\over j!} x_I^k
  \right\}     x_I^j.
 \nonumber
\end{eqnarray}
So 
\begin{eqnarray}
\fl
\hat f^{(i)}_{[f-g]+g}(0) &=& \hat f^{(i)}(0) + g^{(i)}(0) 
\\
\fl
&&
-  i! \; \left[ \sum_I x_I^{i+j} \right]^{-1} \; 
\left\{   \sum_{k=0}^n {g^{(k)}(0)\over k!} \sum_I x_I^{j+k} + 
\sum_{k= n+1}^\infty {g^{(k)}(0)\over k!} \sum_I x_I^{j+k} 
  \right\},   
\nonumber
\end{eqnarray}
whence
\begin{eqnarray}
\hat f^{(i)}_{[f-g]+g}(0) &=& \hat f^{(i)}(0) + g^{(i)}(0) 
-  i! \;    \sum_{k=0}^n    {g^{(k)}(0)\over k!} \left[ \sum_I x_I^{i+j} \right]^{-1} \; 
\left[\sum_I x_I^{j+k}\right]  
\nonumber\\
&& 
- i!  \sum_{k= n+1}^\infty {g^{(k)}(0)\over k!}    \left[ \sum_I x_I^{i+j} \right]^{-1} \sum_I x_I^{j+k}.
\end{eqnarray}
But two of these matrices are simply inverses of each other, so in terms of the Kronecker delta
\begin{eqnarray}
\hat f^{(i)}_{[f-g]+g}(0) &=& \hat f^{(i)}(0) + g^{(i)}(0) 
-  i! \;    \sum_{k=0}^n    {g^{(k)}(0)\over k!} \delta_{ik} 
\nonumber
\\
&& 
- i!  
\sum_{k= n+1}^\infty {g^{(k)}(0)\over k!}    \left[ \sum_I x_I^{i+j} \right]^{-1} \sum_I x_I^{j+k},
\end{eqnarray}
which now leads to significant cancellations
\begin{equation}
\hat f^{(i)}_{[f-g]+g}(0) = \hat f^{(i)}(0) - i!  
\sum_{k= n+1}^\infty {g^{(k)}(0)\over k!}    \left[ \sum_I x_I^{i+j} \right]^{-1} \sum_I x_I^{j+k}.
\end{equation}
This is the best (ignoring convergence issues) that one can do in the general case. Note the formula for the shift of the estimate of the $i$th derivative of $f(x)$ is linear in the derivatives of the function $g(x)$, and that it starts with the $(n+1)$th derivative. Consequently as the order $n$ of the polynomial used to fit the data increases there are fewer terms included in the sum, so the difference between various estimates of the derivatives becomes smaller as more terms are added to the least squares fit.

In the particular situation we discuss in the body of the article
\begin{equation}
f(x) \to \tilde\mu = \ln\left( {d(z)\over  z \hbox{ Mpc}} \right); \quad 
g(x) \to {K\over2}  \ln(1+z); \quad K\in Z;
\end{equation}
or a similar formula in terms of the $y$-redshift. Consequently, from equation (\ref{E:A10}), particularized to our case
\begin{equation}
\fl
\hat {\tilde \mu}^{(i)}_K(0) = \hat {\tilde\mu}^{(i)}(0)  + {K\over2}  [\ln(1+z)]^{(i)}(0) -
{K \; i!\over 2}  
 \left[ \sum_I z_I^{i+j} \right]^{-1} \; \left[\sum_I z_I^{j} \; \ln(1+z_I)\right] .
\end{equation}
Then the ``gap'' between any two adjacent estimates for $\hat {\tilde \mu}^{(i)}_K(0)$ corresponds to taking $\Delta K = 1$ and so
\begin{equation}
\fl
\Delta \hat {\tilde \mu}^{(i)}(0) =  {(-1)^{i-1}\; (i-1)!\over2} -
{i!\over 2}  
 \left[ \sum_I z_I^{i+j} \right]^{-1} \; \left[\sum_I z_I^{j} \; \ln(1+z_I)\right] .
\end{equation}
But then for the particular case $i=1$ which is of most interest to us
\begin{equation}
\hat {\tilde \mu}^{(1)}_K(0) = \hat {\tilde\mu}^{(1)}(0)  + {K\over2} -
{K \over 2}  
 \left[ \sum_I z_I^{i+j} \right]^{-1}_{1j} \; \left[\sum_I z_I^{j} \; \ln(1+z_I)\right] ,
\end{equation}
and
\begin{equation}
\Delta \hat {\tilde \mu}^{(1)}(0) =  {1\over2} -
{1\over 2}  
 \left[ \sum_I z_I^{i+j} \right]^{-1}_{ij} \; \left[\sum_I z_I^{j} \; \ln(1+z_I)\right] .
\end{equation}
By Taylor series expanding the logarithm, and reindexing the terms, this can also be recast as
\begin{equation}
\hat {\tilde \mu}^{(i)}_K(0) = \hat {\tilde\mu}^{(i)}(0)  +
{K \; i!\over 2}  
\sum_{k= n+1}^\infty {(-1)^{k}\over k }    \left[ \sum_I z_I^{i+j} \right]^{-1} \sum_I z_I^{j+k},
\end{equation}
whence
\begin{equation}
\hat {\tilde \mu}^{(1)}_K(0) = \hat {\tilde\mu}^{(1)}(0)  +
{K \over 2}  
\sum_{k= n+1}^\infty {(-1)^{k}\over k }    \left[ \sum_I z_I^{i+j} \right]^{-1}_{1j} \; \sum_I z_I^{j+k},
\end{equation}
and
\begin{equation}
\Delta \hat {\tilde \mu}^{(1)}(0) = 
{1 \over 2}  
\sum_{k= n+1}^\infty {(-1)^{k}\over k }    \left[ \sum_I z_I^{i+j} \right]^{-1}_{1j} \; \sum_I z_I^{j+k},
\end{equation}
(Because of convergence issues, if we work with $z$-redshift these last three formulae make sense only for supernovae datasets where we restrict ourselves to $z_I<1$, working in $y$-redshift no such constraint need be imposed.) 
Now relating this to the modelling ambiguity in $q_0$, we have 
\begin{equation}
\left[\Delta q_0\right]_\mathrm{modelling}   = - 2 \; \Delta \hat {\tilde \mu}^{(1)}(0),
\end{equation}
so that 
\begin{equation}
\left[\Delta q_0\right]_\mathrm{modelling}  = -1 +
 \left[ \sum_I z_I^{i+j} \right]^{-1}_{1j} \; \left[\sum_I z_I^{j} \; \ln(1+z_I)\right] . 
\end{equation}
By Taylor-series expanding the logarithm, modulo convergence issues discussed above, this can also be expressed as:
\begin{equation}
\left[\Delta q_0\right]_\mathrm{modelling}  =
- \sum_{k= n+1}^\infty {(-1)^{k}\over k}    \left[ \sum_I z_I^{i+j} \right]^{-1}_{1j} \;
\left[ \sum_I z_I^{j+k} \right].
\end{equation}
In particular, without further calculation,  these results collectively tell us that the different estimates for $q_0$ will always be evenly spaced, and it suggests that as $n\to\infty$ the differences will become smaller. This is actually what is seen in the data analysis we performed.
\emph{If we were to have a good physics reason for choosing one particular definition of distance as being primary, we would use that for the least squares fit, and the other ways of estimating the derivatives  would be ``biased'' --- but in the current situation we have no physically preferred  ``best'' choice of distance variable.}

\section[\\Combining measurements from different models]{Combining measurements from different models}
\label{A:combine}

\def\Prob{{\mathrm{Prob}}}

Suppose one has a collection of measurements $X_a$, each of which is represented by a random variable $\hat X_a$ with mean $\mu_a = E(\hat X_a)$ and variance $\sigma_a^2 = E( [\hat X_a-\mu_a]^2)$.  How should one then combine these measurements into an overall ``best estimate"?

If we have no good physics reason to reject one of the measurements then the best we can do is to describe the combined measurement process by a random variable $\hat X_{\hat A}$ where $\hat A$ is now a discrete random variable that picks one of the measurement techniques with some probability $p_a$.  More precisely
\begin{equation}
\Prob(\hat A = a) = p_a,
\end{equation}
where the values $p_a$ are for now left arbitrary. Then
\begin{equation}
\mu = E(\hat X_{\hat A}) = \sum_a p_a \; E(\hat X_a) =  \sum_a p_a \; \mu_a,
\end{equation}
and
\begin{equation}
E(\hat X_{\hat A}^2) = \sum_a p_a \; E(\hat X_a)^2 =  \sum_a p_a \; (\sigma^2_a + \mu_a^2).
\end{equation}
But equally well
\begin{equation}
E(\hat X_{\hat A}^2) = \sigma^2 + \mu^2,
\end{equation}
so that overall
\begin{equation}
\mu =  \sum_a p_a \; \mu_a,
\end{equation}
and
\begin{equation}
\sigma^2 = \sum_a p_a \; \sigma^2_a  +    \sum_a p_a \; (\mu_a-\mu)^2.
\end{equation}
This lets us split the overall variance into the contribution from the purely statistical uncertanties on the individual measurements
\begin{equation}
\sigma_\mathrm{statistical} = \sqrt{  \sum_a p_a \; \sigma^2_a },
\end{equation}
plus the ``modelling ambiguity" arising from different ways of modelling the same physics
\begin{equation}
\sigma_\mathrm{modelling} = \sqrt{  \sum_a p_a \; (\mu_a-\mu)^2 }.
\end{equation}
In the particular case we are interested in we have 5 different ways of modelling distance and no particular reason for choosing one definition of measurement over all the others so it is best to take $p_a=1/5$. 

Furthermore in the case of the estimates for the deceleration parameter, all individual estimates have the same statistical uncertainty, and the estimates are equally spaced with a gap $\Delta$:
\begin{equation}
\sigma_a = \sigma_0; \qquad   \mu_a = \mu_P + n \Delta;  \quad n \in\{-2,-1,0,1,2\}.
\end{equation}
Therefore
\begin{equation}
\mu = \mu_P; \qquad \sigma_\mathrm{statistical} = \sigma_0; \qquad \sigma_\mathrm{modelling} = \sqrt{2} \; \Delta.
\end{equation}
For estimates of the jerk, we no longer have the simple equal-spacing rule and equal statistical uncertainties rule,  but there is still no good reason for preferring one distance surrogate over all the others so we still take  $p_a=1/5$ and the estimate obtained from the combined measurements satisfies
\begin{equation}
\fl
\mu = { \sum_a \mu_a\over 5}; \qquad 
\sigma_\mathrm{statistical} = \sqrt{  \sum_a \sigma^2_a\over 5 };
\qquad
\sigma_\mathrm{modelling} = \sqrt{  \sum_a (\mu_a-\mu)^2\over 5 }.
\end{equation}
These formulae are used to calculate the statistical and modelling uncertainties reported in tables \ref{T:q-stat-model}--\ref{T:j-stat-model} and \ref{T:q-all}--\ref{T:j-all} . 
Note that \emph{by definition} the combined purely statistical and modelling uncertainties are to be added in quadrature
\begin{equation}
\sigma = \sqrt{ \sigma_\mathrm{statistical}^2 + \sigma_\mathrm{modelling}^2 }.
\end{equation}
This discussion does not yet deal with the estimated systematic uncertainties (``known unknowns")  or ``historically estimated'' systematic uncertainties (``unknown unknowns'').

\ack

This Research was supported by the Marsden Fund administered by the
Royal Society of New Zealand. CC was also supported by a Victoria University of Wellington Postgraduate scholarship.

\section*{References}
\addcontentsline{toc}{section}{References}


\end{document}